\documentclass[11pt]{article}
\usepackage{epsfig}
\usepackage{amsmath,amssymb}
\usepackage{mathtools}
\usepackage{braket}
\usepackage{hyperref}
\usepackage{authblk}
\usepackage{graphicx}
\usepackage{comment}
\usepackage{soul}
\usepackage{xspace}
\let\OldS\S
\renewcommand{\S}{\OldS\xspace}
\graphicspath{ {./images/} }
\usepackage[section]{placeins}
\usepackage{color}
\DeclareMathOperator{\csch}{csch}

\begin{document}
\date{}

\title{ Holographic timelike entanglement in AdS$_{3}$ Vaidya}

\author[1]{Gaurav Katoch\thanks{gauravitation@gmail.com}}
\author[1]{ Debajyoti Sarkar\thanks{dsarkar@iiti.ac.in}}
\author[1]{Bhim Sen\thanks{bhimsen10496@gmail.com}}

\affil[1]{Department of Physics\\

Indian Institute of Technology Indore\\

 Khandwa Road, Indore, 453552, India}

\maketitle

\begin{abstract}
    Based on the studies of pseudo-entropy in de Sitter, there have been recent proposals for a timelike entanglement in AdS/CFT. In this work, we explore this proposal in the context of a holographic CFT undergoing a global quench. We study various cases in which the timelike intervals are anchored at various boundary times, sometimes straddling the infalling shell. The early and late time behaviours reproduce the known results coming from the pure AdS and the black hole geometry dual to the thermal CFT state respectively. However when the infalling shock straddles the timelike interval, the dynamics drastically differs from how the entanglement entropy evolves. 
\end{abstract}

\newpage
\tableofcontents 
\newpage
\section{Introduction}\label{sec:intro}

The study of strongly coupled field theories had remained a challenging task up until the advent of gauge/gravity or holographic duality \cite{Maldacena:1997re,Gubser:1998bc,Witten:1998qj}. One of the greatest utility of the Anti-de Sitter (AdS)/ conformal field theory (CFT) correspondence is that one may translate the strongly coupled field theory problems entirely into a weakly coupled gravitational problem. Our current work will utilize this duality to investigate some recently studied aspects of the boundary field theory entanglement in a dynamical setting. 
\\

In particular, we will focus on the set-up where the boundary CFT undergoes a global quench. It is well known that the corresponding bulk geometry can be modelled via an infalling shock resulting in an AdS-Vaidya spacetime. To begin with, the system is described by the vacuum state of the CFT, which is subsequently kicked into an excited state following some sudden perturbation. The excited, thermal state of the field theory admits black hole (BH) solutions in asymptotically AdS spacetime as its holographic dual.  A particular manner it can be obtained is by turning on some uniform density operators in the boundary CFT for a very short time and is encoded via a null shell collapsing to form a black hole in the dual holographic picture. On the gravitational side, this Vaidya spacetime dynamically interpolates between pure AdS in the earlier times to a BTZ BH in the far future \cite{Banados:1992wn}.  Such dynamical processes have been exhaustively studied, and they provide us with a holographic prescription of how the field theory approaches thermal equilibrium at a temperature set by the BH temperature. An incomplete list of literature is \cite{Abajo-Arrastia:2010ajo,Ebrahim:2010ra,Albash:2010mv,Balasubramanian:2010ce,Balasubramanian:2011ur,Caceres:2012em,Baron:2012fv,Arefeva:2013kvb,Liu:2013qca,Liu:2013iza}.\\
  
In a parallel development, AdS/CFT has provided us with a productive framework to study several aspects of non-gravitational field theory entanglement in terms of gravitational dynamics. The celebrated Ryu-Takayanagi (RT) conjecture \cite{Ryu:2006bv, Hubeny:2007xt} allows us to compute highly non-trivial quantum entanglement entropies (EE) in terms of far simpler geometric means, where one needs to essentially compute areas of some codimension-two spacelike extremal surfaces (we will generically call them RT surfaces) within the corresponding bulk spacetime. The applications of such a simple proposal is far reaching and vast, which we will not even try to enumerate. However, we will make some comments and observations regarding their potential application to the thermal dynamics mentioned in the above paragraph. After the RT proposal, we realized how boundary entanglement can be an effective probe to study these questions. In fact, RT surfaces can generically traverse the interiors of a BH  geometry for the time-dependent cases \cite{Hubeny:2012ry,Hartman:2013qma,Freivogel:2014lja}, including the  Vaidya spacetime \cite{Buchel:2014gta}.\footnote{Alternatively, various nonlocal operators such as two-point correlators of gauge-invariant operators, Wilson loops (see e.g. \cite{Aparicio:2011zy,Balasubramanian:2011ur,Caceres:2012em,Liu:2013qca} and references therein) etc. have also been employed to investigate how the strongly coupled field theories thermalize.}
\\
 
In a series of even more recent developments, there have been proposals for a holographic framework to account for the entanglement entropy of de Sitter (dS) spacetimes \cite{Narayan:2015vda,Narayan:2016xwq,Jatkar:2017jwz,Narayan:2017xca,Mollabashi:2020yie,Nishioka:2021cxe,Doi:2022iyj,Narayan:2022afv,Doi:2023zaf,Narayan:2023ebn,Parzygnat:2023avh,Narayan:2023zen,Goswami:2024vfl}. Although the precise microscopic construction of dS/CFT duality is still underway, it is a general expectation that Euclidean, non-unitary CFTs dual to the dS spacetime now live on spacelike surfaces at temporal infinities \cite{Strominger:2001pn}. If we now ask questions about quantum entanglements within such set-ups, it results in complex-valued entanglement entropy. Since the reduced density matrices in the dS/CFT turn out to be non-hermitian, this quantity was proposed to be interpreted as pseudo-entropy. It turns out that such complex-valued measures are also obtained if one analytically continues the spacelike holographic EE in AdS to correspond to a timelike subregion at the boundary within AdS/CFT. In \cite{Doi:2022iyj,Narayan:2022afv,Doi:2023zaf,Narayan:2023ebn} this was dubbed timelike entanglement entropy (TEE). In the bulk, the associated quantity can also be alternatively computed in terms of the minimal areas of two disconnected codimension-two spacelike surfaces (being homologous to the timelike region at the boundary), adjoined by a timelike geodesic connecting the other endpoints of the above two spacelike segments. Just like the case for the holographic entanglement entropy (HEE), the areas of these spacelike surfaces contribute to cutoff dependent, universal real part. Whereas the timelike geodesic contributes towards the imaginary part. Together they provide the complex valued TEE. In what follows, when we talk about TEE, we will have such a construction in mind.\footnote{Although it was motivated from the notion of pseudo-entropy, it is perhaps a bit confusing to call this complex-valued quantity timelike \emph{entanglement entropy}. In particular, its statistical origins are not yet properly understood. We will have some comments on this in the conclusion section.}\\

The notion of TEE furnishes an alternate measure alongside its close cousin, HEE. In some cases, it may even prove to be a more sensitive upgrade because, unlike the RT surfaces, the \emph{piece-wise} extremal surfaces corresponding to TEE are able to cross the horizon of the static BHs and approach all the way to the singularity. This attractive feature of TEE enables it to encode exotic features that regular entanglement entropy isn't able to capture so easily (such as its application in sub-horizon bulk reconstruction \cite{Das:2023yyl}, or probing the singularity \cite{Anegawa:2024kdj}).\footnote{This subject is still in its infancy, although a rapidly evolving one. For a very incomplete list of other ongoing works on this emerging field, see e.g. \cite{Heller:2024whi,Jiang:2023loq} etc.} Therefore, it begs the question whether a careful study of this quantity can also be a useful probe for the questions of thermalization. However, we find the answer to this question to be negative, except for the fact that at very early and at very late times our results boil down to what is expected of TEE in the pure AdS and BTZ geometries. In fact, this question of thermalization for a timelike interval was already partially addressed in \cite{Balasubramanian:2012tu}, where a closely related quantity viz.~the expectation value of two timelike separated operators on the boundary dual to AdS Vaidya has been studied. In this work, the authors propose a prescription of Wick rotating the Vaidya geometry (along with two other methods) to obtain a \emph{connected} geodesic. They show that their measure of the associated quantity is a typical probe of field theory thermalization.\footnote{Similar opinions were voiced in \cite{He:2024emd}. According to their classifications, the correct measure of timelike entanglement should be via the \emph{connected, type-C} geodesics, akin to the work of \cite{Balasubramanian:2012tu}. Whereas, here we deal with the disconnected, type-B geodesics along the lines of \cite{Doi:2022iyj, Doi:2023zaf} to make possible connections with the discussions of pseudo-entropy.} However, the way the Wick rotation has been carried out in that work, results in a quite different interpretation compared to the prescription of \cite{Doi:2022iyj, Doi:2023zaf} where the Wick rotated spatial coordinate switches to temporal coordinate and vice versa. For example, in the static case, \cite{Doi:2022iyj, Doi:2023zaf} starts with a thermal or vacuum CFT (dual to appropriate AAdS spacetimes) with a spatial subregion, and Wick rotates it to obtain a vacuum or thermal (respectively) CFT with timelike interval (the Wick rotations are given in section 2.2 of \cite{Doi:2023zaf}, and the results of that section also match with their results in section 2.1). However, the Wick rotation considered in \cite{Balasubramanian:2012tu} is slightly different (page 13 in that paper), which makes the geometry Euclidean, and hence the timelike interval spacelike. Performing computations in such Euclidean geometries, they Wick rotate back their results to obtain the required answer. 
%More recently, \cite{He:2024emd} has proposed a new way of looking at timelike entanglement, along the line of the studies of the timelike correlators. This yields results which are quite different from what we have studied here.
 \\
 
Our main results are contained in section \ref{sec:mainsection}. The different subsections \ref{subsec:beforeqq}, \ref{subsec:straddleearly}, \ref{subsec:straddlelate} and \ref{subsec:afterqq} then study the various arrangements of the boundary intervals with respect to the boundary time and the location of the shock. We have called them cases 1 through 4. In each of these subsections, among other things, we have computed the relevant extremal surfaces, along with their areas and contributions to the TEE. This allows us to understand the dynamics of TEE in our set-up. As a consistency check, we have first reproduced the known results for the timelike subregion living in the asymptotic boundary of pure AdS. A more non-trivial arrangement of boundary interval (case 2, where the interval straddles the shock) reflects a rise in the TEE, hinting that the quenched CFT is in a non-equilibrium state. At very late times (case 4), the (real part of) TEE shows another rise, quickly approaching the asymptotic upper bound set by the TEE of pure BTZ. This is a clear indication that the CFT has thermalized. Finally, we have concluded in section \ref{sec:conclude} with a discussion on some possible future directions.
%\DS{Changed these few lines} 
\\

\textbf{Notes added during the review process:} While our manuscript was under the peer review/ publication process, \cite{Heller:2025kvp,Nunez:2025ppd} appeared on the Arxiv, which argue for a suitably constructed connected surface as being the correct measure for the timelike entanglement entropy for the static states. This is certainly possible, as we currently lack a CFT understanding of entanglement between causally connected states. As of the time of the publication of our present paper, it is not clear to us whether this prescription also yields a reasonable measure of TEE for the Vaidya case. If we consider the connected type-$C$ surface instead of our piecewise ones, some of our preliminary works fail to reproduce a connected curve for TEE as the timelike interval sweeps out the boundary time (we have't discussed these preliminary results here). Moreover, if the above proposal necessarily produces a real valued quantity (which of course makes sense as a measure of `entropy'), the connection of TEE with the dS pseudo-entropy (that was shown to be true in the static case \cite{Doi:2022iyj, Doi:2023zaf}) gets somewhat obscured. It is then also not clear how to recover various limiting cases such as $T_2\to 0$ or $M\to 0$ limits, thereby obtaining the original complex valued results of \cite{Doi:2022iyj, Doi:2023zaf}. However, our results also need to be studied further from a CFT perspective, as it naturally hints at a novel CFT quantity. We thank our anonymous referee for a fervent discussion on these issues. 

%%%%%%%%%%%%%%%%%%%%%%%%%%%%%%%%%%%%%%%%%%%%%%%%%%%%%%%%%%%%%%%%%%%%%%%%%%%
\section{Extremal surfaces for  timelike subregion}\label{sec:mainsection}
%%%%%%%%%%%%%%%%%%%%%%%%%%%%%%%%%%%%%%%%%%%%%%%%%%%%%%%%%%%%%%%%%%%%%%%%%%%

The metric we employ is the AdS$_{2+1}$ Vaidya spacetime 
   \begin{align}
        ds^2=\frac{L^2}{z^2}\left(-f(v,z)\,dv^2-2dvdz+dx^2\right).
    \end{align}
The Vaidya metric describes a null shell composed of tensionless dust. The blackening function is given by $f(v,z)=1-M(v)z^2$, where we have
    \begin{align}
          \begin{array}{cc}
 M(v)\equiv M \Theta(v)=\Bigg\{ & 
\begin{array}{cc}
 M \qquad &\text{for}\qquad v>0 \\
 0 \qquad &\text{for}\qquad  v<0
\\
\end{array}
 \\
\end{array}\nonumber
    \end{align}
\begin{figure}[h]
    \centering
\includegraphics[width=0.5\linewidth]{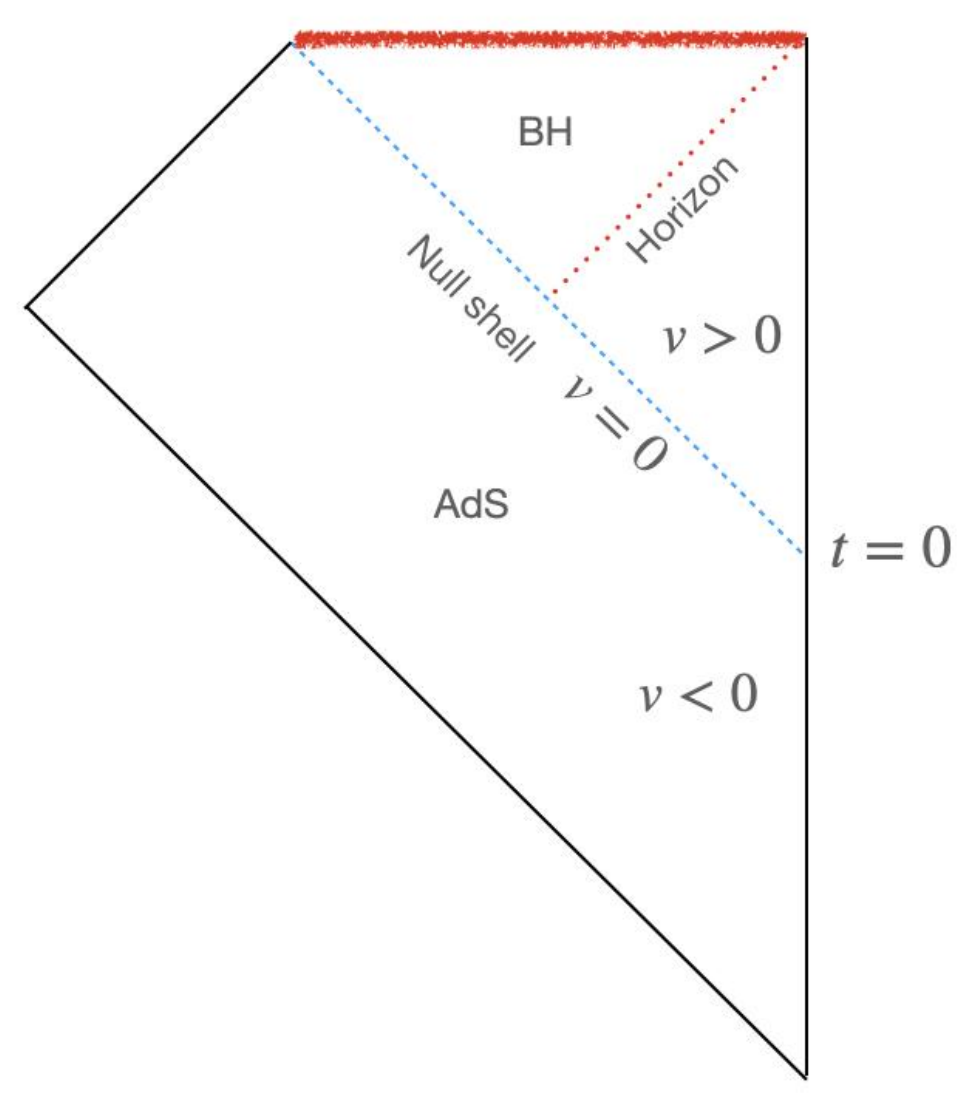}
    \caption{Penrose diagram for AdS Vaidya spacetime due to an infalling null shell at $t=0$ at the boundary (equivalently at $v=0$) represented by the dotted blue line. The region below $v=0$ is the pure AdS spacetime, and the region above corresponds to a black hole. The thick red line represents the singularity, and the red dotted line is the horizon of the black hole. Also, the vertical line on the right represents the boundary time (boundary spatial direction is suppressed), and the null lines on the left are Poincar\'{e} horizons.}
    \label{vaidya}
\end{figure}   
The infalling matter is concentrated with a delta function spike at $v=0$, because we are interested in the zero thickness limit of the collapsing null shell.\footnote{It takes a certain amount of time for the pure AdS state to settle down to BTZ after a quench. It is a standard result of sudden approximation of time dependent perturbation theory, see e.g. chapter XVII, \OldS 8 of \cite{messiah1999quantum}. We will neglect this technicality throughout our analysis.} Here, we have defined the ingoing Eddington-Finkelstein coordinate $v$ as 
         \begin{align}
        v\equiv t-\int^z_0 \frac{dz'}{f(z')}\nonumber\,.
        \end{align}
Henceforth, in this and the following analysis, we have set the scale by plugging AdS radius $L$ to unity. The horizon is located at $z_h=\frac{1}{\sqrt{M}}$  and the inverse temperature of the ($2+1$)-dimensional black hole is given by $\beta=\frac{2\pi}{\sqrt{M}}$. This metric is an uplift from the AdS Poincar\'{e} patch (with coordinates $(t,z,x)$) to a dynamical one due to the perturbation by $M$. For an illustration of this geometry, see figure \ref{vaidya}.\\

In our work, we will be interested in evaluating the quantum extremal surfaces corresponding to timelike subregions anchored at different times at the boundary (but entirely remaining at a constant boundary spatial coordinate $x$). As stated in the introduction, the corresponding surfaces can be inspired by the Wick rotation picture \cite{Doi:2022iyj,Doi:2023zaf} starting with the usual HEE. However, they can also be alternatively obtained by solving for codimension-two extremal surfaces homologous to the boundary time interval. We will take the latter approach in this work and leave the corresponding Wick-rotation story for future work.\\
     
The pullback of the AdS$_3$ Vaidya metric on the codimension-1 surface  $(v(z),x(z))$ is given by
    \begin{align}
        dh^2&=\frac{1}{z^2}\left(-f(z,v)v'(z)^2-2v'(z)+x'(z)^2\right)dz^2\,,\nonumber
    \end{align}
    where $v'(z)=\frac{d v}{dz}$ and $x'(z)=\frac{dx}{dz}$.
    The area functional takes the following form
    \begin{align}
        A&=\int dz\,\frac{\sqrt{-f(v,z)v'(z)^2-2v'(z)+x'^2(z)}}{z}\,. \nonumber
    \end{align}
    As a preliminary check, we observe that $x$ appears as a cyclic coordinate in the area functional. Hence, the corresponding conjugate momentum must be conserved
    \begin{align}
        \frac{x'(z)}{z\sqrt{-f(v,z)v'(z)^2-2v'(z)+x'^2(z)}}=J\,.\nonumber
    \end{align}
    Now, the Neumann boundary condition at the turning point of the $x$ coordinate (since the extremal surface is homologous to the boundary points having fixed $x$ coordinate) dictates $J=0$. In other words, without the loss of generality, the $x$ dynamics of the extremal surface can be dropped out from further analysis. Hence, we will rather analyse the simplified area functional
    \begin{align}\label{Action}
        A=\int dz\,\frac{\sqrt{-f(v,z)v'(z)^2-2v'(z)}}{z}
    \end{align}
    separately in each of the geometric regions of the bulk. The general Euler-Lagrange equation is
    \begin{align}
        2 z v''(z)-f(v,z) \left(z f'(v,z)-2 f(v,z)\right) v'(z)^3-\left(3 z f'(v,z)-6 f(v,z)\right) v'(z)^2+4 v'(z)=0\,.\label{eq:EOM}
    \end{align}
    The  integral of motion is given by 
    \begin{align}\label{eq:intmot}
      \frac{f(v,z) v'(z)+1}{z \sqrt{-v'(z) \left(f(v,z) v'(z)+2\right)}}= P =\text{constant}\,,
    \end{align}
which can be solved for $v'(z)$ to furnish two roots
\begin{align}
  v'(z)= \frac{1}{-f(v,z)-P^2 z^2\pm\sqrt{P^2 z^2 \left(f(v,z)+P^2 z^2\right)}}\,.
\end{align}
A preliminary investigation of the nature of the roots suggests that for both of the roots $v'(z)<0$, and hence the associated curves are spacelike in nature. \\

\textbf{A note on notation}: In the following subsections, we will investigate the behaviour of the solutions, case by case, in the boundary chronology. We have four distinct cases (cases 1 through 4) classified according to the geometrical characteristics of the spacelike extremal surfaces that arise in chronological order. With this in hindsight, we've organised the labelling of the further subsections. We have also uniformly maintained the convention to label the upper and lower roots (and the corresponding solutions) with $+/-$. So, for example, the upper solution pertaining to the case 3 will be referred by $v_{(+3)}(z)$. Similar notations and subscripts have also been used to write down area and entropy functionals, covariant and geodesic distances etc.

%%%%%%%%%%%%%%%%%%%%%%%%%%%%%%%%%%%%%%%%%%%%%%%%%%%%%%%%%%%%%%%%%%%%%%%%%%%    
\subsection{Case 1: Subregion before quench }\label{subsec:beforeqq}
%%%%%%%%%%%%%%%%%%%%%%%%%%%%%%%%%%%%%%%%%%%%%%%%%%%%%%%%%%%%%%%%%%%%%%%%%%%

The first case deals with a timelike CFT  interval located in the far past of the collapsing shell. The spacelike geodesics emanating from the end points of the timelike subregion are restricted to lie entirely in the AdS spacetime at the earliest times.  In this case, the area functional takes the following form
\begin{align}
   A_{AdS}= \int dz\, \frac{\sqrt{-v'(z)^2-2 v'(z)}}{z}\,. \label{AdSArea}
\end{align}
Proceeding along the same arguments made above, we note that the $v$ coordinate is cyclic and hence the conserved $v$-momentum takes the form\footnote{Throughout the paper, we will label the integral of motion that appears in \eqref{eq:intmot} as $p$ or $P$, depending on whether we are in the AdS or in the BH segment.}
\begin{align}
    \frac{v'(z)+1}{z \sqrt{-v'(z) \left(v'(z)+2\right)}}=p\,.\nonumber
\end{align}
The above algebraic equation can be solved for $v'(z)$, giving us two solutions
\begin{align}
    v'_{-}(z)=\frac{1}{-1-p^2 z^2+\sqrt{p^2 z^2 \left(1+p^2 z^2\right)}} \quad \text{and} \label{AdSv1}
    \end{align}
    \begin{align}
 v'_{+}(z)= \frac{1}{-1-p^2 z^2-\sqrt{p^2 z^2 \left(1+p^2 z^2\right)}}\,,  \label{AdSv2}
\end{align}
furnishing us with the two roots.

%%%%%%%%%%%%%%%%%%%%%%%%%%%%%%%%%%%%%%%%%%%%%%%%%%%%%%%%%%%%%%%%%%%%%%%%%%%
\subsubsection{Lower AdS branch $(T_1<0)$  }
%%%%%%%%%%%%%%%%%%%%%%%%%%%%%%%%%%%%%%%%%%%%%%%%%%%%%%%%%%%%%%%%%%%%%%%%%%%

\begin{figure}[h!tbp]
    \centering
\includegraphics[width=0.6\linewidth, height=0.4\linewidth]{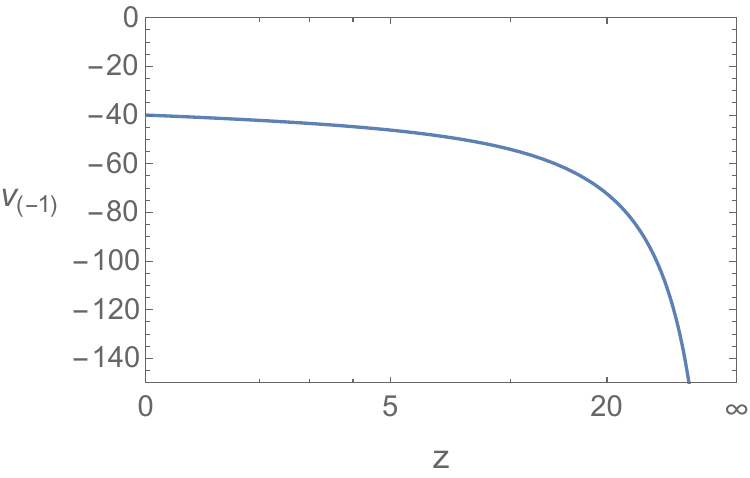}
    \caption{$v_{(-1)}$ as a function of the bulk radial coordinate $z$ for the interval size $T_2-T_1=20$ and $T_1=-40$. $v_{(-1)}$ reaches negative infinity as $z\to \infty$.}
    \label{vminus1}
\end{figure}

We'll first discuss the extremal surface anchored to the lower endpoint of the subregion at boundary time $v=t=T_1$. Because in case 1, we expect our results to boil down to the results of \cite{Doi:2022iyj,Doi:2023zaf}, we expect the spacelike surface from $T_1$ to hit the lower Poincar\'{e} horizon. It turns out that for such a demand, in order to find the corresponding $v(z)$, we have to integrate \eqref{AdSv1}. In that case, we have (using the notations discussed above)
\begin{align}
    v_{(-1)}(z)=-z-\frac{\sqrt{p^4 z^2+p^2}}{p^2}+c_1\,.\nonumber
\end{align}
Upon demanding the correct boundary behaviour $v_{(-1)}(0)=T_1$, it fixes $p=\pm\frac{1}{(c_1 - T_1)}=p_{\pm}$ with $c_1 > T_1$. With both choices of  $p=p_{\pm}$, we find that the spacelike branch of the extremal surface emanating from the lower point $T_1$ is given by 
\begin{align}
    v_{(-1)}(z)=-z-\sqrt{(c_1-T_1)^2+z^2}+c_1 \,.\label{AdS1l}
\end{align}
It can be seen that deep in the bulk, this branch indeed hits the lower edge $v=-\infty$ of the Poincar\'{e} horizon, as shown in figure \ref{vminus1} (plotted for the choice of $c_1=\frac{T_1+T_2}{2}$, as motivated below).\\ 

\begin{figure}[h]
    \centering
\includegraphics[width=0.5\linewidth]{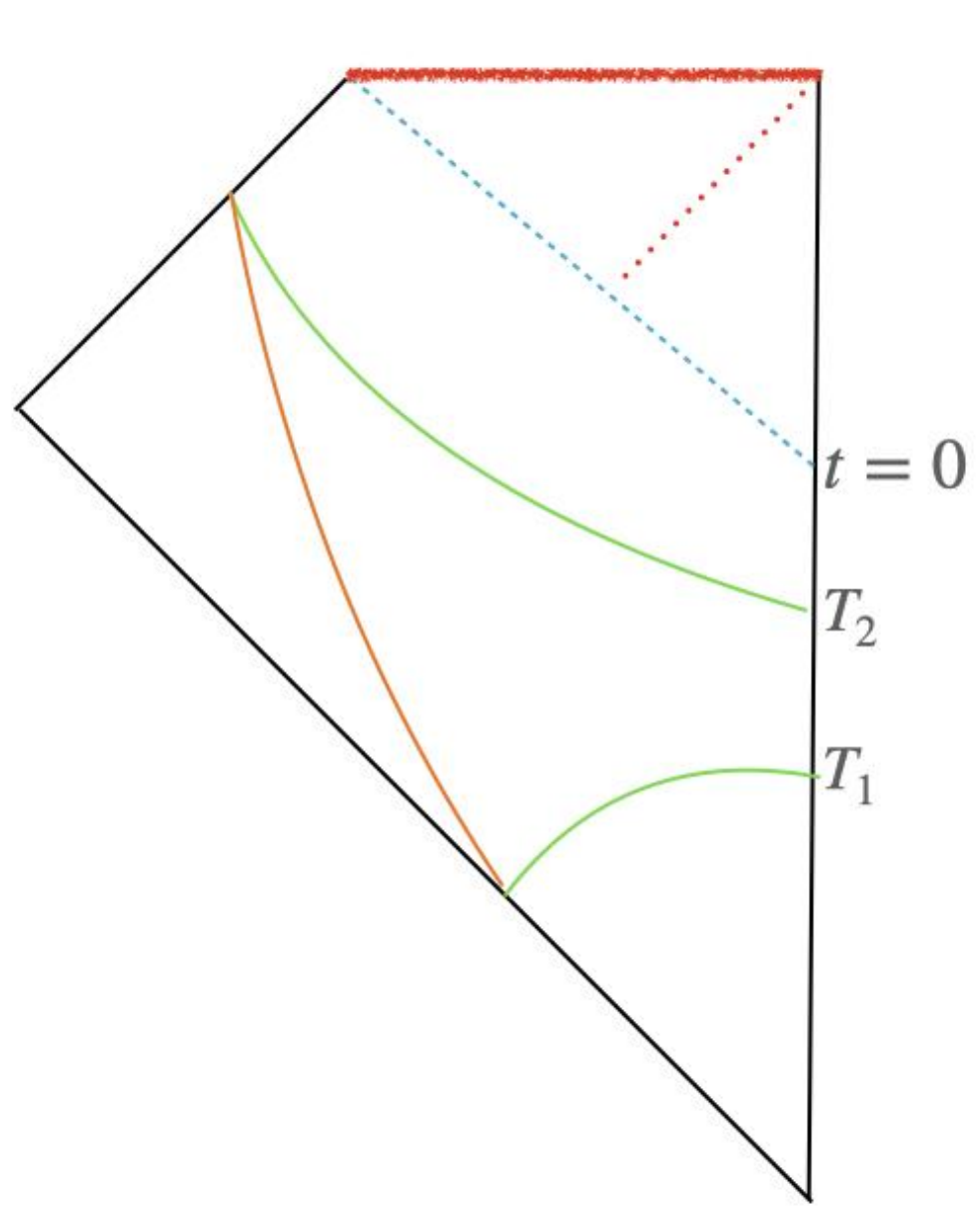}
    \caption{Schematic diagram of earlier time behaviour when the subregion is in pure AdS.}
    \label{AdS}
\end{figure}

On plugging (\ref{AdS1l}) into (\ref{AdSArea}), the extremal area for the lower branch of the spacelike curve turns out to be 
\begin{equation}\label{AdS-1}
	A_{(-1)}=\int_{\epsilon}^{\infty} dz\, \frac{\sqrt{-(v'_{(-1)}(z))^2-2v'_{(-1)}}}{z}=\sinh ^{-1}\left(\frac{c_1-T_1}{\epsilon}\right)\simeq \log\left(\frac{2(c_1-T_1)}{\epsilon}\right)\,,
\end{equation}
where in writing the last equation above, we have neglected all the $\mathcal{O}(\epsilon^2)$ pieces and higher.
The quantity $\epsilon$ above denotes the usual UV cutoff of the boundary theory, which also arises in the computations of the HEE. Below, we will perform analogous computations for the upper spacelike branch and the resulting timelike segment. However, a schematic diagram of the surfaces is already given in figure \ref{AdS}. In this diagram and in the others to follow, the green curves will always correspond to the spacelike segments and the orange curve to the timelike one. 

%%%%%%%%%%%%%%%%%%%%%%%%%%%%%%%%%%%%%%%%%%%%%%%%%%%%%%%%%%%%%%%%%%%%%%%%%%%
\subsubsection{Upper AdS branch $(T_2<0)$ }
%%%%%%%%%%%%%%%%%%%%%%%%%%%%%%%%%%%%%%%%%%%%%%%%%%%%%%%%%%%%%%%%%%%%%%%%%%%

In this case, we consider the other extremal surface anchored to the other boundary time $v=t=T_2$. In other words, our subregion size will always be $T=T_2-T_1$ (in our case 1, $T_1, T_2<0$ with $T_2>T_1$). Now this time, consistency with \cite{Doi:2022iyj,Doi:2023zaf} demands that we need to integrate (\ref{AdSv2}) to obtain the general solution 
\begin{align}
    v_{(+1)}(z)=-z+\frac{\sqrt{p^4 z^2+p^2}}{p^2}+c_2\,.\nonumber
\end{align}
Once again, requiring the correct boundary behaviour $v_{+1}(0)=T_2$, fixes $p=\pm\frac{1}{(c_2-T_2)}$ with $c_2<T_2$. Once again, for both values of $p$, we obtain the upper AdS root as
\begin{align}
    v_{(+1)}(z)=-z+\sqrt{(c_2-T_2)^2+z^2}+c_2 \label{AdS+1u}\,.
\end{align}
Indeed, this upper AdS root now meets the upper edge of the Poincar\'{e} horizon at some finite value $v(\infty)=c_2$. The constraint such as $c_2<T_2$ can be understood physically as the spacelike nature of the corresponding curves. 
\begin{figure}[h]
    \centering
\includegraphics[width=0.6\linewidth,height=0.4\linewidth]{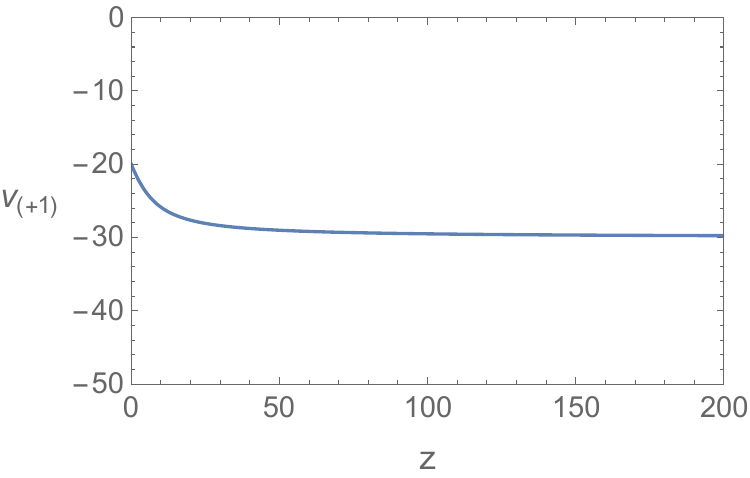}
    \caption{$v_{(+1)}$ as a function of the bulk radial coordinate $z$ for the interval size $T_2-T_1=20$ and $T_1=-40$. The curve goes all the way to $z\to\infty$, with the correct asymptotic values for $v_{(+1)}$ at $z=0$ and $z=\infty$.}
    \label{vplus1}
\end{figure}
Once again, a plot for this upper root with the choice $c_2=\frac{T_1+T_2}{2}$ is given in figure \ref{vplus1}.\footnote{Note that both our computations for $v_{(\pm 1)}$ was motivated from the results of \cite{Doi:2022iyj,Doi:2023zaf}. Without this motivation, we can have extremal surfaces both ending at either the upper or the lower Poincar\'{e} horizon. Such a possibility has recently been discussed in \cite{Anegawa:2024kdj}. Similar ambiguities may also arise when we discuss the other cases, but we will always be motivated by making contacts with the known results of \cite{Doi:2022iyj,Doi:2023zaf}, which is in-turn consistent with the Wick rotation picture as mentioned in the introduction. It will be interesting to explore these various possibilities and their effects on the dynamics of TEE.}
\\
 
Combining (\ref{AdS+1u}) and (\ref{AdSArea}), we obtain the total extremal area of the spacelike geodesics to be (in future, we will always omit the $\mathcal{O}(\epsilon^2)$ terms)
\begin{equation}
	A_{(+1)}=-\sinh ^{-1}\left(\frac{c_2-T_2}{\epsilon}\right)= \log\left(\frac{2(T_2-c_2)}{\epsilon}\right)+\mathcal{O}(\epsilon^2)\,. \label{AdS+1}
\end{equation}
Since $p$ is the integral of motion \eqref{eq:intmot}, common for both the upper and the lower roots, this facilitates eliminating one integration constant from our equations giving $c_1=-c_2+T_1+T_2$. Moreover, motivated by the variational treatment (see appendix \ref{var} for further details), we were able to fix $c=c_1=c_2=\frac{T_1+T_2}{2}$ for the current case 1. For this particular choice, the TEE expressions are rightfully just functions of the subregion length $(T_2-T_1)=T$, giving a precise agreement with the pre-existing results in the literature \cite{Doi:2022iyj,Doi:2023zaf}. In particular, we have the total length of the spacelike portions of the extremal curves as (simply the sum of (\ref{AdS+1}) and (\ref{AdS-1}))
\begin{equation}
    A_{(+1)}+A_{(-1)}=A_{1}=2\log\left(\frac{T_2-T_1}{\epsilon}\right)\,.\label{AdS1}
\end{equation}
In fact it turns out that also for case 4, the natural relation between the integration constants is given by $c_1=-c_2+T_1+T_2$. Hence we will choose the above values of $c_1$ and $c_2$ in all the following cases. This consistent choice keeps the transitions of extremal surfaces smooth between one case to the other, and makes the final expression of the timelike entanglement entropy to depend only on $T$. 

%%%%%%%%%%%%%%%%%%%%%%%%%%%%%%%%%%%%%%%%%%%%%%%%%%%%%%%%%%%%%%%%%%%%%%%%%%%
\subsubsection{Timelike segment}
%%%%%%%%%%%%%%%%%%%%%%%%%%%%%%%%%%%%%%%%%%%%%%%%%%%%%%%%%%%%%%%%%%%%%%%%%%%

To account for the imaginary part of the pseudo-entropy coming from the Wick rotation picture \cite{Doi:2022iyj, Doi:2023zaf},  and to obtain one connected surface homologous to the boundary interval, we have to join the non-boundary anchored (free) endpoints of the spacelike surfaces with a timelike geodesic. The length of the timelike geodesic $\Delta$ can be easily evaluated by utilising the AdS covariant distance $\sigma$ between the two bulk points located at coordinates $(z_1,v_1)$ and $(z_2,v_2)$ \cite{Hamilton:2006fh}. In particular, they are related by $\Delta=\cosh^{-1}(\sigma(z_2,v_2|z_1,v_1))$.
In our case, as obtained in the previous subsections, the free endpoint corresponding to the lower root $v_{(-1)}(z)$ is located at the Poincar\'{e} horizon $(\Lambda,-2\Lambda)$ ($\Lambda$ is a large $z$, IR regulator). Whereas the free endpoint of the upper root $v_{+1}(z)$ goes to $(\Lambda,c_2)$. The associated AdS covariant distance is therefore given by (below the subscript in $\sigma_1$ denotes case 1)
\begin{equation}
    \sigma_{1}=\lim_{\Lambda\to \infty}\left(\frac{z_2^2+z_1^2-(v_2-v_1+(z_2-z_1))^2}{2z_1z_2}\right)\to -1\,.
\end{equation}
Thus, the geodesic length of the timelike curve turns out to be
\begin{align}
    \Delta_{1}=\cosh^{-1}(-1)=i\pi.\label{AdSt1}
\end{align}
Note that the time-like geodesic is purely imaginary and has no dependence upon the boundary time. This is analogous to what happens for the static cases, as discussed in \cite{Doi:2022iyj, Doi:2023zaf}.
\\

%%%%%%%%%%%%%%%%%%%%%%%%%%%%%%%%%%%%%%%%%%%%%%%%%%%%%%%%%%%%%%%%%%%%%%%%%%%
\subsubsection{Holographic TEE}
%%%%%%%%%%%%%%%%%%%%%%%%%%%%%%%%%%%%%%%%%%%%%%%%%%%%%%%%%%%%%%%%%%%%%%%%%%%

\begin{figure}
    \centering
    \includegraphics[width=0.5\linewidth]{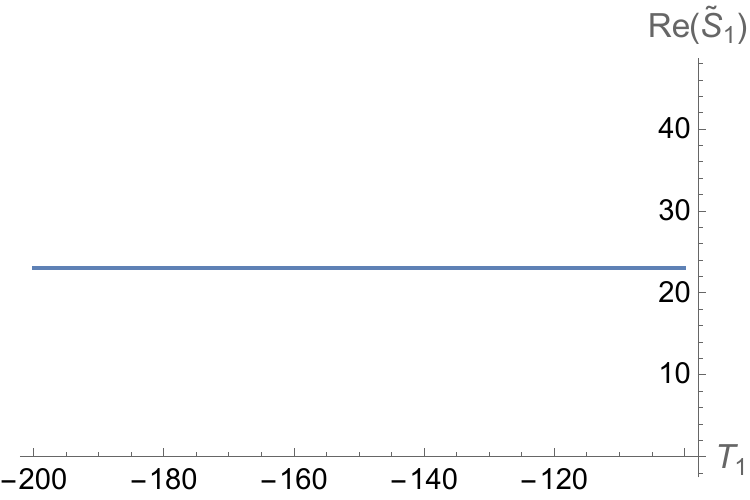}
    \caption{Real part of holographic TEE for case 1. The plot uses a subregion size $T_2-T_1=100$ and $\epsilon=0.001$.}
    \label{HTEE1}
\end{figure}

Combining  (\ref{AdS1}) and (\ref{AdSt1}) and dividing by the appropriate factor of the Planck area (length for $d=2$), we obtain the following familiar result for TEE
\begin{align}
    \tilde{S}_{1}&=\frac{c}{3}\log\left(\frac{T_2-T_1}{\epsilon}\right)+\frac{c}{6}\pi i.
\end{align}
This result was derived in \cite{Doi:2022iyj, Doi:2023zaf} within pure AdS/CFT. For our case 1, such a match is expected because the associated extremal surfaces of the given subregion are not sensitive to the shock present in the future. The appearance of the imaginary part can be traced back to the fact that unlike the usual RT curve (which is a connected spacelike curve in its entirety), here we get two disconnected spacelike curves. Therefore, we require a timelike geodesic to reduce the minimal surface into a connected piece. As mentioned before, this measure for TEE can also be derived using the Wick rotation picture.\footnote{In fact, as illustrated in \cite{Das:2023yyl}, one can also obtain these results by defining a timelike modular Hamiltonian by Wick rotating the usual spacelike modular Hamiltonian.} Finally, we have plotted the corresponding entropy in figure \ref{HTEE1}, which shows its dependence as the function of the boundary time $T_1$. Since the holographic TEE only depends on the subregion size and the UV cutoff $\epsilon$ (taken at a fixed value), it remains constant until $T_2<0$.

%%%%%%%%%%%%%%%%%%%%%%%%%%%%%%%%%%%%%%%%%%%%%%%%%%%%%%%%%%%%%%%%%%%%%%%%%%%
\subsection{Case 2: Subregion straddling the shock - early time behaviour}\label{subsec:straddleearly}
%%%%%%%%%%%%%%%%%%%%%%%%%%%%%%%%%%%%%%%%%%%%%%%%%%%%%%%%%%%%%%%%%%%%%%%%%%%

The next case we will consider is when the subregion of size $T$ has evolved upwards and is now located at $T_1<0$, $T_2=T_1+T>0$, such that $|T_2|\leq |T_1|$. In this second case, the upper spacelike surface begins its journey from the  BTZ region, but later crosses the shock and enters the AdS side. However, the lower spacelike surface is still entirely in the AdS region. 

%%%%%%%%%%%%%%%%%%%%%%%%%%%%%%%%%%%%%%%%%%%%%%%%%%%%%%%%%%%%%%%%%%%%%%%%%%%
\subsubsection{Lower AdS branch ($T_1<0$)}
%%%%%%%%%%%%%%%%%%%%%%%%%%%%%%%%%%%%%%%%%%%%%%%%%%%%%%%%%%%%%%%%%%%%%%%%%%%

In this case, the analysis of the lower root, which lies in the AdS region, is the same as before and we have rewritten it below with the appropriate notation
\begin{align}
    v_{(-2)}(z)=-z-\sqrt{(c_1-T_1)^2+z^2}+c_1 .\label{AdS2l}
\end{align}
Once again, this root ends up at the lower  Poincar\`e horizon i.e. at $v_{(-2)}(\infty)=-\infty$ leading to a similar expression for the extremal length (as in (\ref{AdS-1}))
\begin{align}
 A_{(-2)}=  \log\left(\frac{T_2-T_1}{\epsilon}\right).\label{TEE-2}
\end{align}
\begin{figure}[h]
    \centering
\includegraphics[width=0.5\linewidth]{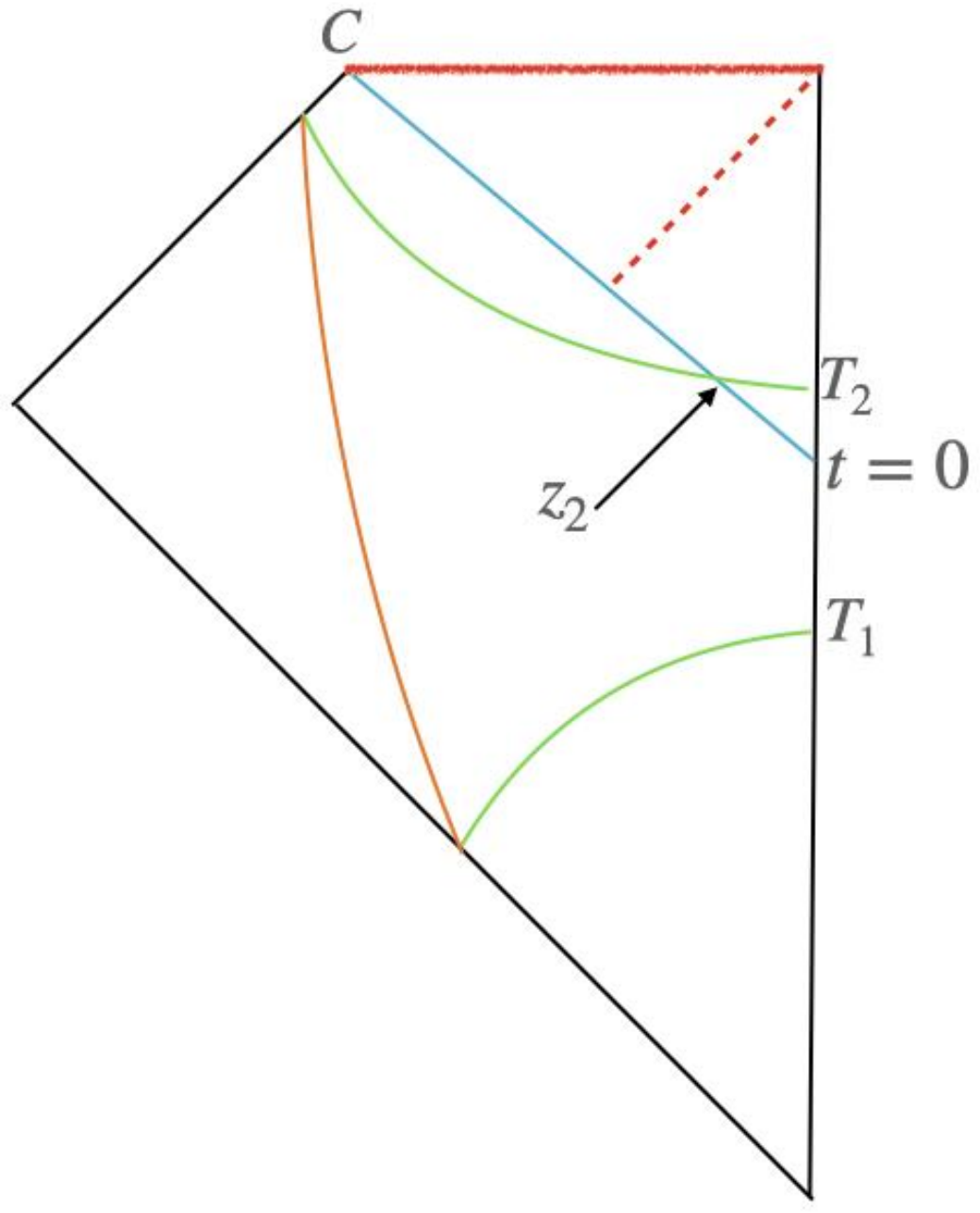}
    \caption{Subregion starts to enter the black hole region with $T_2>0$ and $T_1<0$. As $T_2\to|T_1|$, $z_2$ approaches the point $C$.}
    \label{AdSBTZ1}
\end{figure}

%%%%%%%%%%%%%%%%%%%%%%%%%%%%%%%%%%%%%%%%%%%%%%%%%%%%%%%%%%%%%%%%%%%%%%%%%%%
\subsubsection{Upper BTZ branch ($T_2>0$)}
%%%%%%%%%%%%%%%%%%%%%%%%%%%%%%%%%%%%%%%%%%%%%%%%%%%%%%%%%%%%%%%%%%%%%%%%%%%

The upper branch however, behaves more non-trivially. For a certain range of the values of positive $T_2$, it starts from the BH side but comes out to the AdS side at a value $z=z_2$ (referred to here as the crossing point). A schematic diagram is given in figure \ref{AdSBTZ1}. Given the crossing, one would require to compute the corresponding junction conditions. We have derived the condition in appendix \ref{app:jc} (see e.g.~\cite{Balasubramanian:2011ur,Liu:2013iza} for a derivation of the same in the context of RT surfaces in AdS-Vaidya). The main equation takes the form 
%\DS{Added some new lines here}.
\begin{equation}\label{eq:jc}
	f^B(v,z) v'^B(z)=v'^A(z)\,,
\end{equation}
where the superscripts $A$ and $B$ denotes the AdS and BTZ regions respectively. However, we will notice that for the upper surface (i.e.~the surface attached to $T_2$), the junction conditions are not well-defined for the entire range where case 2 is valid (see appendix \ref{app:case2alt} for a brief discussion of this point). Therefore, we will resort to the gluing condition \eqref{eq:gc} for the treatment of this part.\\

In what follows, we will solve for the behaviour of the upper surface explicitly, and evaluate the crossing point.
Rewriting the area functional (\ref{Action}) explicitly for the BTZ case, we have
\begin{align}\label{BTZArea}
    A_{BTZ}=\int dz\, \frac{\sqrt{-(1-Mz^2)v'(z)^2-2v'(z)}}{z}\,.
\end{align}
This time, the solution of \eqref{eq:EOM} leads to the following integral of motion 
   \begin{align}
      \frac{(1-Mz^2) v'(z)+1}{z \sqrt{-v'(z) \left( v'(z)(1-Mz^2) +2\right)}}= P\,.\nonumber
    \end{align}
Once again, this algebraic equation has the following two roots
\begin{align}
    v'_{-}(z)&=\frac{1}{-(1-Mz^2)-P^2 z^2+Pz\sqrt{ \left((1-Mz^2)+P^2 z^2)\right)}}\label{BTZv-}\,,
\end{align}
\begin{align}
    v'_{+}(z)&=\frac{1}{-(1-Mz^2)-P^2 z^2-Pz\sqrt{ \left((1-Mz^2)+P^2 z^2)\right)}}\label{BTZv+}\,.
\end{align}
Given our set-up, we will only be working with (\ref{BTZv+}) which corresponds to the upper spacelike branch originating in the BTZ region. We obtain\footnote{Note our slightly modified notation of $v_{(+2)}^B$, and later $v_{(+2)}^A$. They denote the upper root of this second case with support both in the BH (B) and the AdS (A) region. Later on, we will also use such superscripts in denoting the geodesic lengths $\Delta$ in the respective parts of the spacetime.} 
\begin{align}
    v^B _{(+2)}(z)=\frac{1}{\sqrt{M}}\left(\tanh ^{-1}\left(\frac{\sqrt{M} \sqrt{P^2  \left(z^2 \left(P^2-M\right)+1\right)}}{P^2 }\right)-\tanh ^{-1}\left(\sqrt{M} z\right)\right)+c'_2\,. \label{BTZv2+}
\end{align}
The requirement of obeying the consistent boundary condition $v_{(+2)}(0)=T_2$ forces the integration constant to be
\begin{align}\label{eq:BHsidemom}
 P=\pm\sqrt{M} \coth \left(\sqrt{M} \left(T_2-c'_{2}\right)\right)\,.    
\end{align}
Furthermore, at the crossing point, $v_{(+2)}(z_2)=0$, so that 
\begin{align}
  z_2=  \frac{\csch \left(c'_{2} \sqrt{M}\right) \left(\cosh \left(\sqrt{M} \left(c'_{2}-T_2\right)\right)-\cosh \left(c'_{2} \sqrt{M}\right)\right)}{\sqrt{M}}\,. \label{crossing1}
\end{align}
\begin{figure}
    \centering
\includegraphics[width=0.5\linewidth]{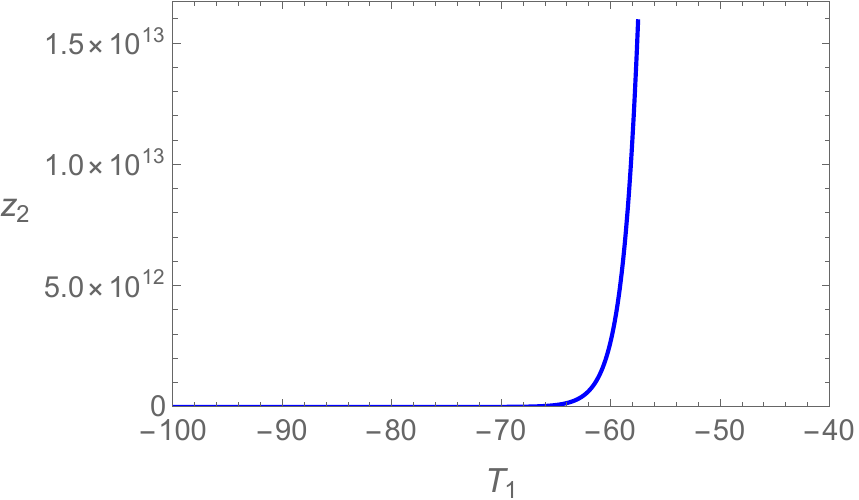}
    \caption{$z_2$ increases monotonically as a function of the boundary time $T_1$, and goes upto $\infty$ as  $|T_1|=T_2$. The above figure is plotted for $M=0.5$, and subregion size $T_2-T_1=100$.}
    \label{z2}
\end{figure}
It can be seen that the expression for the crossing point $z_2$ is a monotonically increasing function of the boundary time $T_1$ as shown in figure \ref{z2}, and it approaches the Poincar\'e horizon (i.e.~$z_2\to\infty$) the moment  $T_2=|T_1|$. Thereafter, all the upper spacelike curves fall into the singularity, something that we will see happen from the next cases onwards. \\

As the extremal surface crosses over into the pure AdS region, its continuation is dictated by integrating the root (\ref{AdSv2}) (we can alternatively derive this by putting $M\to 0$ limit in \eqref{BTZv2+}, which also implies that $c_2=c'_2$ below)
\begin{align}
    v^A_{(+2)}(z)=-z+\frac{\sqrt{p^4 z^2+p^2}}{p^2}+c_2\,.
\end{align}
For consistency, we demand that $v_{(+2)}(z_2)=0$ which forces $ p=\pm\frac{1}{\sqrt{c_2 \left(c_2-2 z_2\right)}}$ with $c_2<0$.\\
\begin{figure}[h]
    \centering
\includegraphics[width=0.5\linewidth]{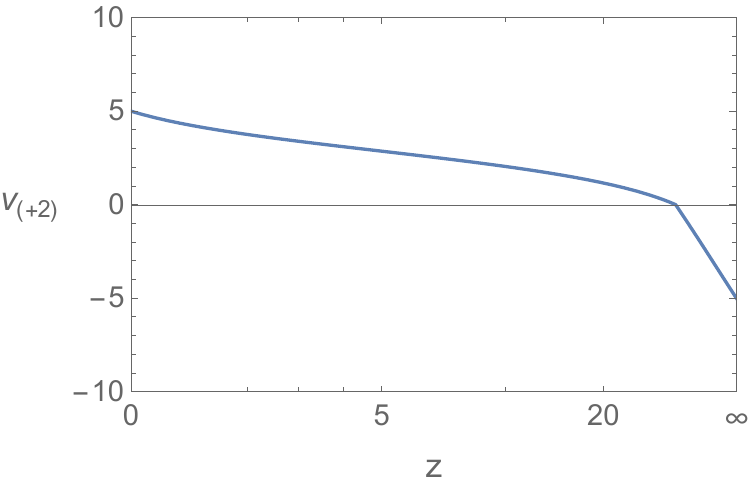}
    \caption{$v_{(+2)}$ as a function of the bulk radial coordinate $z$ for the interval size $T_2-T_1=20$, $T_1=-15$ and $M=0.5$. It can seen that $v_{(+2)}$ takes correct boundary and asymptotic values at $z=0$ and $z=\infty$.}
    \label{vplus2}
\end{figure}

Putting all of this together, we can write the pure AdS continuation of the upper BTZ solution as
\begin{align}
    v^A_{(+2)}(z)=-z+c_2+\sqrt{-2 c_2 z_2+c_2^2+z^2}\,. \label{AdSv2+}
\end{align}
Furthermore, it can be checked that this solution ends up at the upper edge of the Poincar\'{e} horizon, i.e. $v_{(+2)}(\infty)=c_2$. 
Combining the two expressions  (\ref{BTZv2+}) and (\ref{AdSv2+}), the spacelike solution lying partially in both geometries can be concisely written as
\begin{align}\label{eq:gc}
          \begin{array}{cc}
  v_{(+2)}(z)=\Bigg\{ & 
\begin{array}{cc}
   v^B _{(+2)}(z) \qquad &\text{for}\qquad v>0\,, \\
  v^A_{(+2)}(z) \qquad &\text{for}\qquad  v<0\,.
\\
\end{array}
 \\
\end{array}
    \end{align}
%\begin{align}
%    v_{(+2)}(z)=(1-\Theta(z-z_2))  v^B _{(+2)}(z)+\Theta(z-z_2) v^A_{(+2)}(z)\,.
%\end{align}
An exact plot is given in figure \ref{vplus2} for the choices of $c_2=c'_2=\frac{T_1+T_2}{2}$.\footnote{It should be noted that for numerical limitations of Mathematica, we can't plot $v_+(z)$ for too large subsystem size (i.e. for a very large $T$) whenever it is in the BH region.}
\\

Therefore, the length of the extremal surface is given by
\begin{align}
   A_{(+2)}&=A_{(+2)}^B+A_{(+2)}^{A}\nonumber\\
   &= \int_{\epsilon}^{z_2}dz\, \frac{\sqrt{-(1-Mz^2)(v'_{(+2)}(z))^2-2v'_{(+2)}(z)}}{z}+\int_{z_2}^{\infty} dz\, \frac{\sqrt{-(v'_{(+2)}(z))^2-2v'_{(+2)}}}{z}\,. \nonumber
\end{align}
When evaluated separately (keeping track of which part is the AdS root and which one is for BH), we obtain
\begin{align}
    A_{(+2)}^{A}=\sinh ^{-1}\left(\frac{\sqrt{(T_1+T_2) \left((T_1+T_2)-4 z_2\right)}}{2 z_2}\right) \quad \text{and}
\end{align}
\begin{align}
    A_{(+2)}^{B}=\log \left(\frac{2 \sinh \left(\frac{1}{2} \sqrt{M} \left(T_1-T_2\right)\right)}{\epsilon \sqrt{M}}\right)+\sinh ^{-1}\left(\frac{\sinh \left(\frac{1}{2} \sqrt{M} \left(T_2-T_1\right)\right)}{z_2 \sqrt{M}}\right)\,.\nonumber
\end{align}
Adding them together, we have
\begin{align}
    A_{(+2)}=\log \left(\frac{2 \sinh \left(\frac{1}{2} \sqrt{M} \left(T_2-T_1\right)\right)}{\epsilon \sqrt{M}}\right)+\sinh^{-1}\left(\frac{T_2-T_1}{2 \epsilon}\right)+\Phi_2(T_1,T_2,M)\label{TEE+2}\,,
    \end{align}
    where we have repackaged the cumbersome  cutoff independent finite expression as 
  {\small{  \begin{align}
   &\Phi_2(T_1,T_2)=-\sinh ^{-1}\left(\frac{1}{2 \sqrt{\sinh ^2\left(\frac{\sqrt{M} T_1}{2}\right) \sinh ^2\left(\frac{\sqrt{M} T_2}{2}\right) \text{csch}^2\left(\frac{1}{2} \sqrt{M} \left(T_1-T_2\right)\right) \text{csch}^2\left(\frac{1}{2} \sqrt{M} \left(T_1+T_2\right)\right)}}\right)\nonumber\\
    &+\sinh ^{-1}\Bigg[\frac{1}{4} \sqrt{M} \sinh \left(\frac{1}{2} \sqrt{M} \left(T_1+T_2\right)\right) \csch\left(\frac{\sqrt{M} T_1}{2}\right)\csch\left(\frac{\sqrt{M} T_2}{2}\right)\times \nonumber\\ &\sqrt{\frac{\left(T_1+T_2\right) \left(\sqrt{M} \left(T_1+T_2\right)-4 \coth \left(\frac{1}{2} \sqrt{M} \left(T_1+T_2\right)\right)+4 \cosh \left(\frac{1}{2} \sqrt{M} \left(T_1-T_2\right)\right) \csch\left(\frac{1}{2} \sqrt{M} \left(T_1+T_2\right)\right)\right)}{\sqrt{M}}}\Bigg]\,.
\end{align}}}

%%%%%%%%%%%%%%%%%%%%%%%%%%%%%%%%%%%%%%%%%%%%%%%%%%%%%%%%%%%%%%%%%%%%%%%%%%%
\subsubsection{Timelike segment}
%%%%%%%%%%%%%%%%%%%%%%%%%%%%%%%%%%%%%%%%%%%%%%%%%%%%%%%%%%%%%%%%%%%%%%%%%%%

In this case, as both the spacelike solutions end up at the upper and lower edges of the Poincar\'{e} horizon, the timelike geodesic is obtained by joining those endpoints. This situation is reminiscent of the one already encountered in the previous section. Therefore, we can follow a similar route and read off the geodesic length from (\ref{AdSt1}) to be  $\Delta_{2}=\cosh^{-1}(-1)=i\pi.$ 

%%%%%%%%%%%%%%%%%%%%%%%%%%%%%%%%%%%%%%%%%%%%%%%%%%%%%%%%%%%%%%%%%%%%%%%%%%%
\subsubsection{Holographic TEE}
%%%%%%%%%%%%%%%%%%%%%%%%%%%%%%%%%%%%%%%%%%%%%%%%%%%%%%%%%%%%%%%%%%%%%%%%%%%

\begin{figure}[h]
    \centering
    \includegraphics[width=0.5\linewidth]{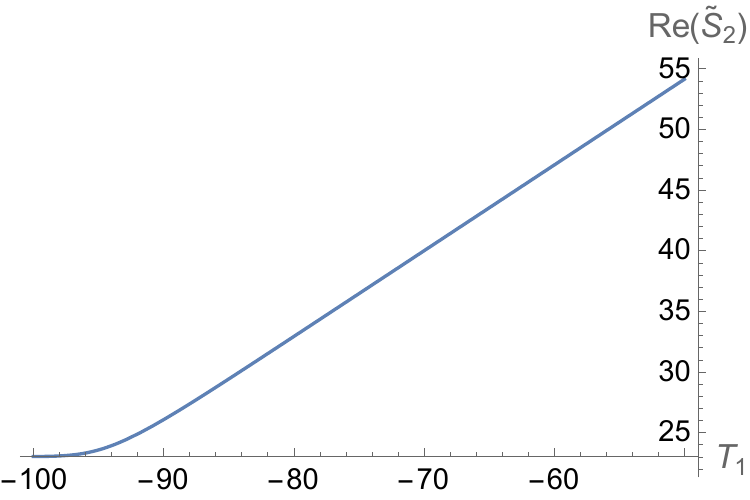}
    \caption{Real part of holographic TEE for case 2. The plot uses a subregion size $T_2-T_1=100$, $M=0.5$ and $\epsilon=0.001$.}
    \label{TEE2}
\end{figure}
Holographic TEE for case 2 turns out to be the sum of (\ref{TEE-2}), (\ref{TEE+2}) and $\Delta_{2}$, and can be written as 
\begin{align}
    \tilde{S}_{2}=\frac{c}{6}\log \left(\frac{\beta }{\pi \epsilon }\sinh \left(\frac{\pi}{\beta}\left(T_2-T_1\right)\right)\right)+\frac{c}{6}\log\left(\frac{T_2-T_1}{ \epsilon}\right)+\frac{c}{6}
\Phi_2(T_1,T_2,\beta)+\frac{c}{6}\pi i\,.
\end{align}
The expression for $\tilde{S}_{2}$ is separated into the sum of two universal divergent pieces characteristic of the TEE (and HEE) and a finite piece independent of the UV cutoff. Its real part displays a rise in TEE, which is solely attributable to the finite piece $\Phi_2(T_1,T_2,\beta)$ in the expression (see figure \ref{TEE2}). Here, the appearance of inverse temperature hints at the fact that $\beta$ also sets the scale for the rate at which equilibration happens. If we study the above $\tilde{S}_{2}$ as a function of the average time $T_{avg}=\frac{T_1+T_2}{2}$ or $T_1$, then one can see the rate of this above-mentioned rise to be linear in the corresponding variable. This linear rise persists for a time of order $T/2$, after which the case 3 takes over. This interesting feature is characteristic of the fact that the extremal surface is in the process of probing the full BTZ region as it moves from the UV region to the IR region. The growth sustains as long as the spacelike extremal surface crosses the null shell before hitting the confluence of the upper Poincar\'e horizon and the black hole singularity denoted by point $C$ in figure \ref{AdSBTZ1}. The dual CFT state, in this case, is evolving towards the state of thermal equilibrium from a far-from-equilibrium state. The first and the second terms above are the expected pieces from the respective static situations in BH and AdS \cite{Doi:2022iyj,Doi:2023zaf}.\\
%Interestingly, this is reminiscent of what happens for spatial subregions, where the linear rise persists for a time or order half of the subregion size, after which it saturates (as we will see shortly, we also have a saturating TEE for case 3) \cite{Calabrese:2005in}

It will be beneficial in passing to remark that the analogous studies of thermalization in the past have also encountered a linear growth associated to entanglement entropy in the earlier times \cite{Liu:2013qca,Balasubramanian:2012tu}. Since we are dealing with a timelike subregion, in particular, only with the areas of the spacelike parts of the extremal surfaces and not with a spacelike subregion (in other words, the associated quantities are fundamentally different), we will not delve into a detailed comparison and contrast with that study. 

% \DS{Changed this paragraph a bit}

%%%%%%%%%%%%%%%%%%%%%%%%%%%%%%%%%%%%%%%%%%%%%%%%%%%%%%%%%%%%%%%%%%%%%%%%%%%
\subsection{Case 3: Subregion straddling the shock - intermediate time behaviour }\label{subsec:straddlelate}
%%%%%%%%%%%%%%%%%%%%%%%%%%%%%%%%%%%%%%%%%%%%%%%%%%%%%%%%%%%%%%%%%%%%%%%%%%%

\begin{figure}[h]
    \centering
\includegraphics[width=0.5\linewidth]{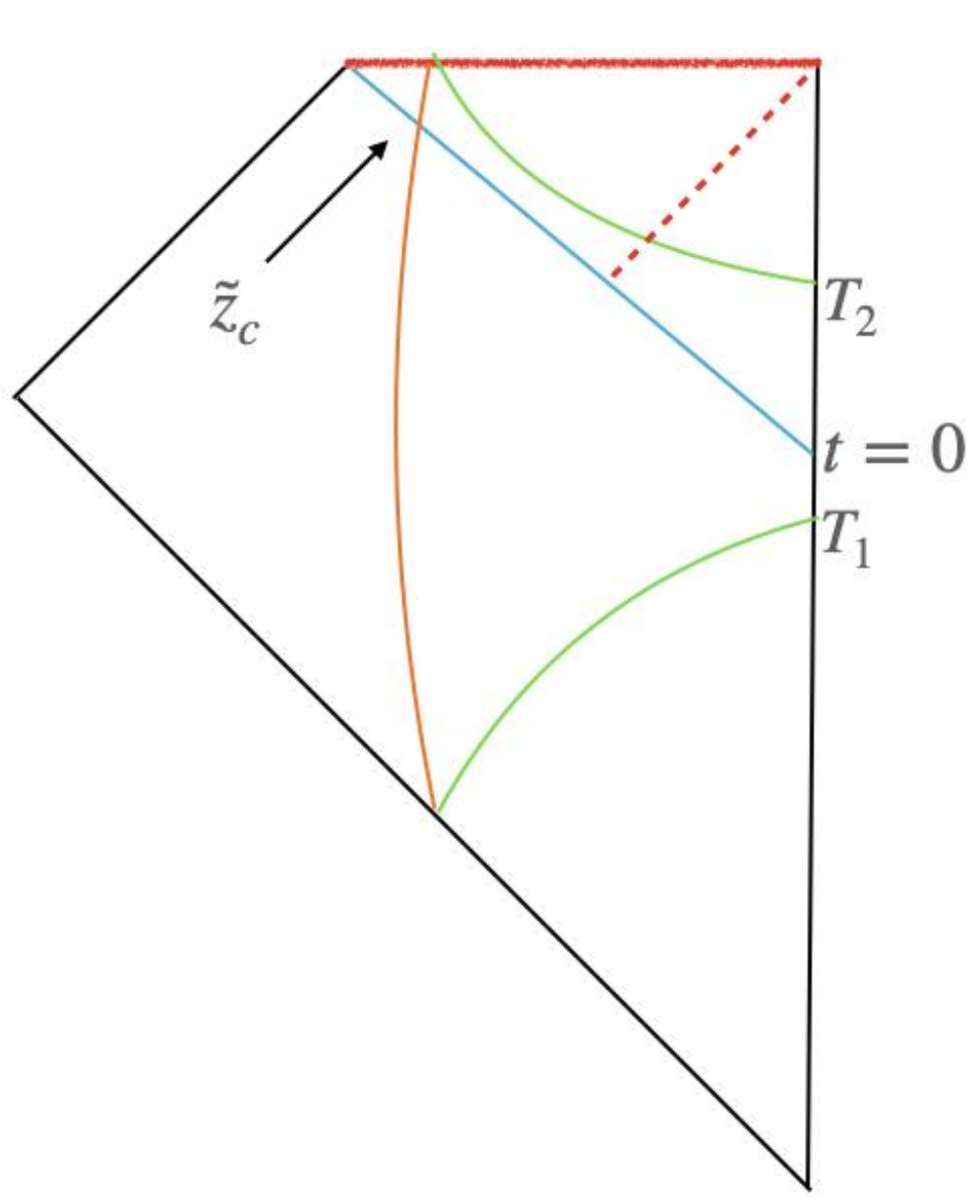}
    \caption{The upper RT surface is now completely in the BH region and goes up to the singularity.}
    \label{fig:AdSBTZ}
\end{figure}

In the previous section we noticed that for $T_1<0$, $T_2>0$ and for $|T_1|>|T_2|$, the upper branch crosses the shell at $z=z_2$ and enters the AdS region. Moreover, $z_2$ goes to the confluence point between the horizon and the singularity in the limit $|T_2|\to|T_1|$. In this section, we will be analysing the case when $|T_1|<|T_2|$, with the boundary region still straddling the shock. In this scenario, the lower root stays purely in the AdS region, ending at the Poincar\'{e} horizon. Whereas the upper branch stays entirely in the BH regime. The timelike segment however crosses the shell at a point $z=\tilde{z}_c$. A schematic diagram is in figure \ref{fig:AdSBTZ}.

%%%%%%%%%%%%%%%%%%%%%%%%%%%%%%%%%%%%%%%%%%%%%%%%%%%%%%%%%%%%%%%%%%%%%%%%%%%
\subsubsection{Lower AdS branch ($T_1<0$)}
%%%%%%%%%%%%%%%%%%%%%%%%%%%%%%%%%%%%%%%%%%%%%%%%%%%%%%%%%%%%%%%%%%%%%%%%%%%

The analysis of the lower AdS branch is unaltered from all the former cases, resulting in the expression similar to  (\ref{AdS-1}) for the length of the lower AdS root 
\begin{align}
    A_{(-3)}= 
 \log\left(\frac{T_2-T_1}{\epsilon}\right)\,.\label{A-3}
\end{align}

%%%%%%%%%%%%%%%%%%%%%%%%%%%%%%%%%%%%%%%%%%%%%%%%%%%%%%%%%%%%%%%%%%%%%%%%%%%
\subsubsection{Upper BTZ branch ($T_2>0$)}
%%%%%%%%%%%%%%%%%%%%%%%%%%%%%%%%%%%%%%%%%%%%%%%%%%%%%%%%%%%%%%%%%%%%%%%%%%%

The upper BTZ solution is obtained by integrating the root (\ref{BTZv+}) subjected to the boundary condition $v_{(+3)}(0)=T_2$. The spacelike minimal curves are now able to penetrate the black hole horizon quite generically and probe the black hole interior. The resulting expression is given by 

\begin{figure}[h]
    \centering
\includegraphics[width=0.5\linewidth]{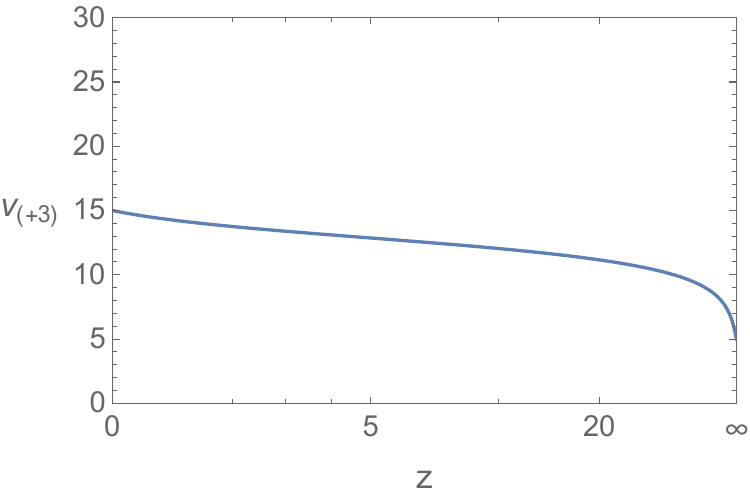}
    \caption{$v_{(+3)}$ as a function of the bulk radial coordinate $z$ for the interval size $T_2-T_1=20$, $T_1=-5$ and $M=0.5$. It can be seen that $v_{(+3)}$ takes correct boundary and asymptotic values for $z=0$ and $z=\infty$.}
    \label{vplus3}
\end{figure}

{\small{\begin{align}
   v_{(+3)}(z)=\frac{-1}{\sqrt{M}} \tanh ^{-1}\left(\tanh \left(\sqrt{M} \left(c_{2}-T_2\right)\right) \sqrt{M z^2 \csch^2\left(\sqrt{M} \left(c_{2}-T_2\right)\right)+1}\right)+\tanh ^{-1}\left(\sqrt{M} z\right)+c_2\label{BTZ3v+}\,,
\end{align} }}
and has been plotted in figure \ref{vplus3} for the choice of $c_2=\frac{T_1+T_2}{2}$.\\

The above equation is similar to (\ref{BTZv2+}) where we have substituted the value of (compare with \eqref{eq:BHsidemom})
\[
P=\pm\sqrt{M} \coth \left(\sqrt{M} \left(T_{2}-c_2\right)\right)\,.
\]
This value for $P$ can be obtained by using the boundary homology condition  $v_{(+3)}(0)=T_2$.  For this curve, the solution can be continued all the way to $z=\infty$ where it reaches the BH singularity for $c_2>0$ at a finite value of $v$.
When we plug in (\ref{BTZ3v+}) into (\ref{BTZArea}), we obtain 
\begin{align}
A_{(+3)}=    \log \left(\frac{2 \sinh \left(\frac{1}{2} \sqrt{M} \left(T_2-T_1\right)\right)}{\epsilon \sqrt{M}}\right)\label{A+3}\,.
\end{align}
This turns out to be the same as what one obtains for the pure BTZ case of  \cite{Doi:2022iyj,Doi:2023zaf}.

%%%%%%%%%%%%%%%%%%%%%%%%%%%%%%%%%%%%%%%%%%%%%%%%%%%%%%%%%%%%%%%%%%%%%%%%%%%
\subsubsection{Timelike segment }
%%%%%%%%%%%%%%%%%%%%%%%%%%%%%%%%%%%%%%%%%%%%%%%%%%%%%%%%%%%%%%%%%%%%%%%%%%%

In this case, the timelike curve emanates from the point $(z,v)=(\infty,-\infty)$ on the lower edge of the Poincar\'{e} horizon, crosses the null shell at the point $(\tilde{z}_c,0)$ and charters off into the BTZ region eventually meeting the point $(\infty, c_2)$ on the BTZ singularity. Introducing the IR regulator $\Lambda$, we find that the geodesic length of the part of the curve that lies in the pure AdS portion of the Vaidya spacetime (bounded by points  $(\Lambda,-2\Lambda)$ and $(\tilde{z}_c,0)$) is given by 
\begin{equation}
  \Delta^A_{3}= \lim_{\Lambda\to \infty} \cosh ^{-1}\left(\frac{\Lambda ^2+\tilde{z}_c^2-(\tilde{z}_c-(-\Lambda ))^2}{2 \Lambda  x}\right)=i\pi\,.
\end{equation}
The other portion of the curve in the BTZ region is bounded by the endpoints $(\tilde{z}_c,0)$ and $(\infty, c_2)$. To evaluate this geodesic distance, we employ the Rindler decomposition of the BTZ in the future Rindler wedge, as given in equation (37) of \cite{Hamilton:2006fh}. Note that the null collapsing shell is now located at the Rindler horizon $r_+=\sqrt{M}$, and the endpoints of the timelike curve are labelled by their $(r,\tilde{v})$ values in the Rindler coordinates.\footnote{$\tilde{v}$ labels some finite value of the ingoing Eddington-Finkelstein coordinate for the future Rindler wedge.} In terms of these, the coordinates of the lower and the upper endpoints of this portion of the timelike segment are given by $(r_+,0)$ and $(0,\tilde{v})$, respectively. Due to the lack of angular dependence, we are left with the simplified-looking expression for the AdS covariant distance
\begin{equation}
    \sigma(0,c_2|\sqrt{M},0)=\lim_{\substack{r'\to r_+\\ r\to0}}\left(\frac{rr'}{r_+^2}\mp\left(1-\frac{r^2}{r_+^2}\right)^{1/2}\left(1-\frac{r'^2}{r_+^2}\right)^{1/2}\sinh(\tilde{t}-\tilde{t}')\right)=0\,.
\end{equation}
Recalling the relation between the geodesic distance $\Delta$ with $\sigma$, it is easy to see that the geodesic distance turns out to be 
\begin{align}
\Delta^B_{3}&=\frac{i\pi }{2}\,.\nonumber
\end{align}
Putting them together, we, therefore, have the total length of the timelike curve as
\begin{align}
    \Delta_{3}
    &=\frac{3i\pi}{2}\,.\label{T3}
\end{align}
We note that the timelike segment has purely an imaginary part and shows no dependence on the crossing point. Even though it straddles across the null surface and sees both the geometries, it fails to capture the dynamical content. It would be interesting to further understand the origin of such a distinct imaginary piece from the boundary perspective.

%%%%%%%%%%%%%%%%%%%%%%%%%%%%%%%%%%%%%%%%%%%%%%%%%%%%%%%%%%%%%%%%%%%%%%%%%%%
\subsubsection{Holographic TEE }
%%%%%%%%%%%%%%%%%%%%%%%%%%%%%%%%%%%%%%%%%%%%%%%%%%%%%%%%%%%%%%%%%%%%%%%%%%%

\begin{figure}[h]
    \centering
    \includegraphics[width=0.5\linewidth]{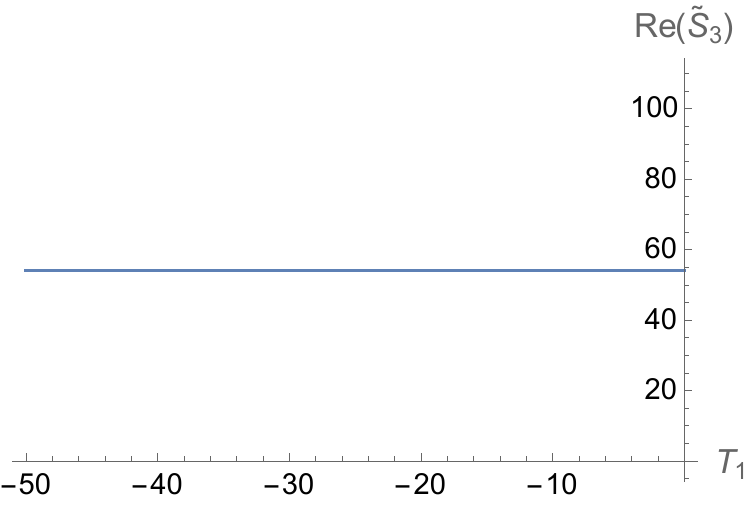}
    \caption{Real part of holographic TEE for case 3. The plot uses a subregion size $T_2-T_1=100$, $M=0.5$ and $\epsilon=0.001$.  }
    \label{fig:TEE3}
\end{figure}
Combining the contributions from all the segments (\ref{A-3}), (\ref{A+3}) and (\ref{T3}), we get the following expression for holographic TEE for case 3  
\begin{align}
    \tilde{S}_{3}=\frac{c}{6}\log \left(\frac{\beta }{\pi \epsilon }\sinh \left(\frac{\pi}{\beta}\left(T_2-T_1\right)\right)\right)+\frac{c}{6}\log \left(\frac{T_2-T_1}{\epsilon}\right)+\frac{c}{4}i\pi\,. \label{TEE3}
\end{align}
As we can see, its general form is only a function of the subregion length and hence is independent of the boundary time $T_1$. This behaviour is reminiscent of the static case,  and does not capture the dynamical features of the geometry.\\

It is interesting to note that even though, in this case, we are probing a dynamical geometry and the timelike segment of our extremal curve crosses the shell, the final TEE is static and does not show any time dependence as shown in Figure \ref{fig:TEE3}. Instead our result contains individual contributions that one would expect to get separately from the pure black hole and pure AdS geometry. The existence of static behaviour at the intermediate time is antithetical with how a CFT is supposed to thermalize \cite{Balasubramanian:2010ce,Balasubramanian:2012tu, Liu:2013iza, Liu:2013qca}. Hence, existence of the static region indicates that perhaps TEE is not an optimal quantity to probe thermalization. 
% \textit{This makes sense in the hindsight because unlike other probes of thermalization viz two point correlators and entanglement entropy that are defined at same time, the dof participating in TEE are spread over time.} \DS{Is this last line concrete? If not, can we write something concrete?} }

%%%%%%%%%%%%%%%%%%%%%%%%%%%%%%%%%%%%%%%%%%%%%%%%%%%%%%%%%%%%%%%%%%%%%%%%%%%
\subsection{Case 4: Subregion after quench - late time behaviour }\label{subsec:afterqq}
%%%%%%%%%%%%%%%%%%%%%%%%%%%%%%%%%%%%%%%%%%%%%%%%%%%%%%%%%%%%%%%%%%%%%%%%%%%

\begin{figure}[h]
    \centering
\includegraphics[width=0.5\linewidth]{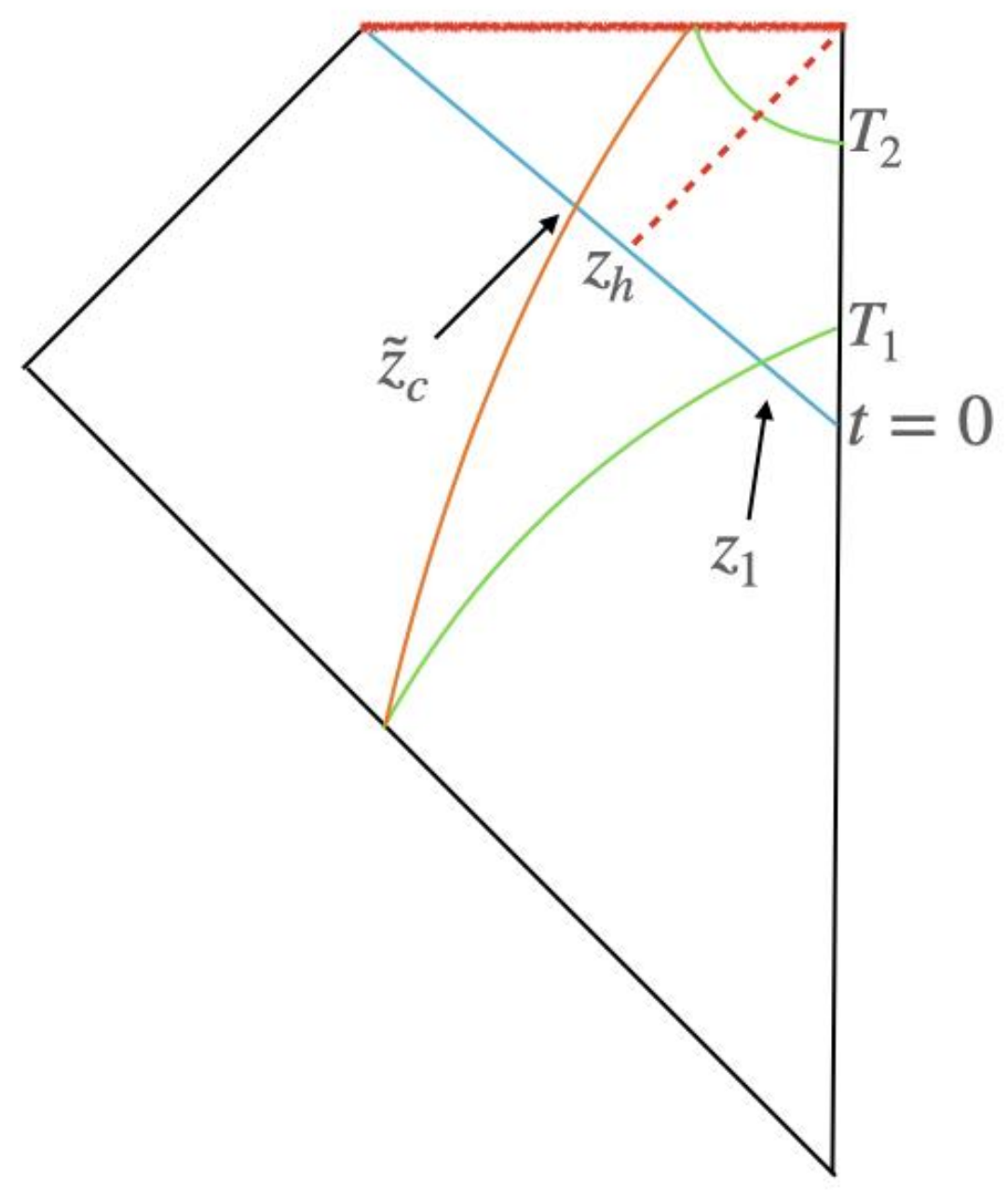}
    \caption{Lower RT surface starts entering into the AdS region. It intersects the null shell at $z=z_1$. The maximum value that $z_1$ attains is $z_h$.}
    \label{AdSBTZ}
\end{figure}

The fourth case arises at late times, when both endpoints are at positive values of $t$ (i.e. $0< T_1$). Therefore, in this setup, the subregion is deep into the boundary region asymptotic to the black hole. In this case, the upper spacelike extremal surface passes through the horizon and ultimately meets its fate at the BH singularity. However, the lower spacelike branch, as well as the timelike segment, crosses the shell over to the AdS side of the geometry and ends up at the Poincar\'e horizon. The resulting spacetime diagram schematically looks like figure \ref{AdSBTZ}.

%%%%%%%%%%%%%%%%%%%%%%%%%%%%%%%%%%%%%%%%%%%%%%%%%%%%%%%%%%%%%%%%%%%%%%%%%%%
\subsubsection{Upper BTZ branch}
%%%%%%%%%%%%%%%%%%%%%%%%%%%%%%%%%%%%%%%%%%%%%%%%%%%%%%%%%%%%%%%%%%%%%%%%%%%

The behaviour of the upper solution for the fourth case remains unaltered from the behaviour already encountered for the case 3. Similar to that case, it is destined to hit the BTZ singularity at some finite $v$-value of $c_2$.  The expression for the spacelike solution $v_{(+4)}(z)$ is similar in form to $v_{(+3)}(z)$, and is given by (\ref{BTZ3v+}). This results in the minimal area (that can simply be read off from $A_{(+3)}$ in (\ref{A+3})) 
\begin{align}
A_{(+4)}=    \log \left(\frac{2 \sinh \left(\sqrt{M} \left(T_2-T_1\right)\right)}{\epsilon \sqrt{M}}\right)\,.\label{A+4}
\end{align}

%%%%%%%%%%%%%%%%%%%%%%%%%%%%%%%%%%%%%%%%%%%%%%%%%%%%%%%%%%%%%%%%%%%%%%%%%%%
\subsubsection{Lower BTZ branch}
%%%%%%%%%%%%%%%%%%%%%%%%%%%%%%%%%%%%%%%%%%%%%%%%%%%%%%%%%%%%%%%%%%%%%%%%%%%

This lower spacelike solution for the fourth case is more interesting than the upper solution because it is destined to end up at the lower Poincar\'e horizon upon crossing the shell. In hindsight, it is also required to do so in order to maintain a valid causal character for the adjoining timelike segment. \\

The lower spacelike curve starts off at $T_1>0$ in the BTZ region and can be obtained by solving (\ref{BTZv-}). The correct boundary behaviour ($v_{(-4)}(0)=T_1$) can be recovered by taking $P=\pm\sqrt{M} \coth \left(\sqrt{M} \left(c_{1}'-T_1\right)\right)$. We therefore obtain

\begin{figure}
    \centering
    \includegraphics[width=0.5\linewidth]{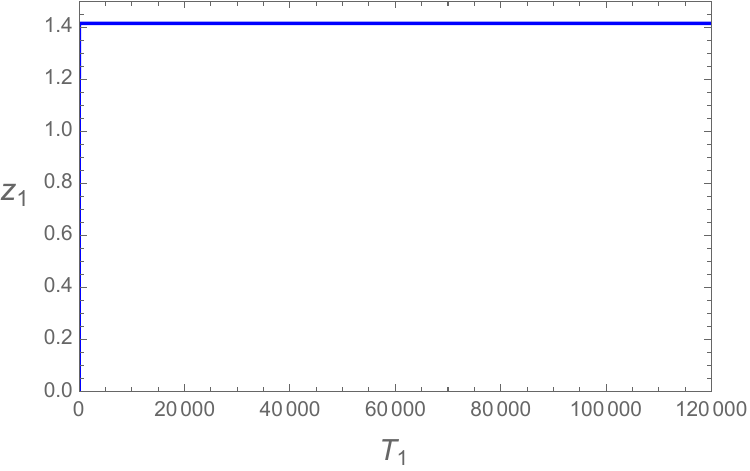}
    \caption{$z_1$ grows for small value of $T_1$, eventually saturating to $z=z_h$. The above figure is plotted for $M=0.5$, which translates to $z_h=1.4$. }
    \label{z1}
\end{figure}

{\small{\begin{align}
   v^B_{(-4)}(z)= c'_{1}-\frac{\tanh ^{-1}\left(\tanh \left(\sqrt{M} \left(c'_{1}-T_1\right)\right) \sqrt{M z^2 \csch^2\left(\sqrt{M} \left(c'_{1}-T_1\right)\right)+1}\right)+\tanh ^{-1}\left(\sqrt{M} z\right)}{\sqrt{M}}\,.\label{BTZv4}
\end{align}}}
The zero of this solution occurs at the crossing point 
\begin{align}
  z_1=  \frac{\text{csch}\left(c'_{1} \sqrt{M}\right) \left(\cosh \left(\sqrt{M} \left(c'_{1}-T_1\right)\right)-\cosh \left(c'_{1} \sqrt{M}\right)\right)}{\sqrt{M}}\label{crossing2}\,.
\end{align}
Since for this case $c'_1=\frac{T_1+T_2}{2}$ can be taken to be arbitrarily large and positive, it can be seen that $z_1$ asymptotes to the value  $\frac{1}{\sqrt{M}}$ for all the later boundary times $T_1\gg 0$ (see figure \ref{z1}). This is nothing but the value of the $z$ coordinate at the horizon.
As a result, unlike the upper surface in the previous case 2, the lower surface only asymptotically approaches the black hole horizon but is unable to see through it. This is in fact a necessity, if we are to obtain a timelike segment at all. However, this feature is also helpful for us to implement the junction condition consistently.\\

\begin{figure}[h]
    \centering
\includegraphics[width=0.5\linewidth]{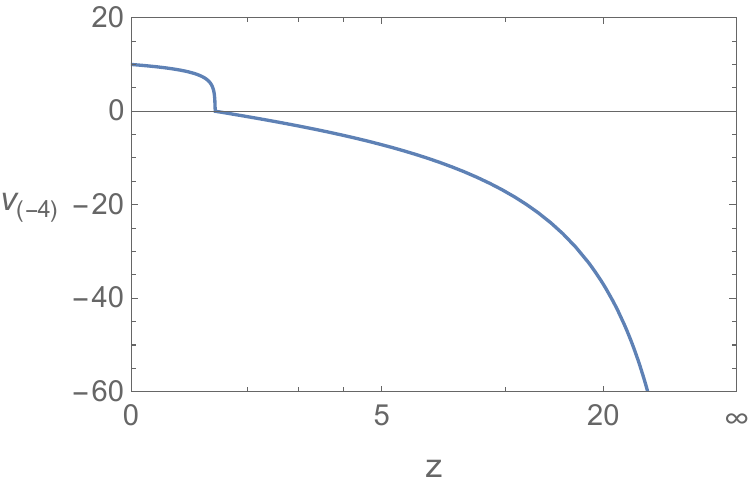}
    \caption{$v_{(-4)}$ as a function of the bulk radial coordinate $z$ for the interval size $T_2-T_1=20$, $T_1=10$ and $M=0.5$. It can be seen that $v_{(-4)}$ takes correct boundary and asymptotic values for $z=0$ and $z=\infty$.}
    \label{vminus4}
\end{figure}

After crossing the shell, the lower surface smoothly transitions over to the pure AdS region by following the junction conditions
\begin{align}
    q=\pm \frac{P}{\sqrt{1-M z_1^2}}\,,
\end{align}
which can in-turn be derived from the original condition \eqref{eq:jc} or \eqref{junction}. This gives the corresponding solution for the AdS part as 
\begin{align}
    v^{A}_{(-4)}(z)=-z+z_1-\frac{\sqrt{-M z_1^2+P^2 z^2+1}}{P}+\frac{\sqrt{z_1^2 \left(P^2-M\right)+1}}{P}\,.\label{AdSv4}
\end{align}
%before eventually ending up at the lower edge of the Poincar\'{e} horizon ($v_{(-4)}(z\to\infty)\to-\infty$). 
Combining (\ref{BTZv4}) and (\ref{AdSv4}), the resulting spacelike solution can be concisely written as (see figure \ref{vminus4} for a plot of this surface) 
\begin{align}
          \begin{array}{cc}
  v_{(-4)}(z)=\Bigg\{ & 
\begin{array}{cc}
   v^A _{(-4)}(z) \qquad &\text{for}\qquad v>0\,, \\
  v^B_{(-4)}(z) \qquad &\text{for}\qquad  v<0\,.
\\
\end{array}
 \\
\end{array}
    \end{align}
%    \textcolor{red}{This is $z_1$ not $z_2$.}
%\begin{align}
%    v_{(-4)}(z)=(1-\Theta(z-z_2))v^B_{(-4)}(z)+\Theta(z-z_2)v^A_{(-4)}(z)\,.\nonumber
%\end{align}

The length of the extremal curve can be written as the separate sum
\begin{align}
   A_{(-4)}&=A_{(-4)}^B+A_{(-4)}^{A}\nonumber\\
   &= \int_{\epsilon}^{z_1}dz\, \frac{\sqrt{-(1-Mz^2)(v'_{(-4)}(z))^2-2v'_{(-4)}(z)}}{z}+\int_{z_1}^{\infty} dz\, \frac{\sqrt{-(v'_{(-4)}(z))^2-2v'_{(-4)}}}{z}\,. \label{A-4}
\end{align}
Evaluating the corresponding integrals, we get for the BTZ part 
\begin{align}
    A_{(-4)}^B=    \log \left(\frac{2 \sinh \left(\frac{1}{2} \sqrt{M} \left(T_2-T_1\right)\right)}{\epsilon \sqrt{M}}\right)-  \sinh^{-1}\left(\frac{\sinh\left(\frac{1}{2} \sqrt{M} \left(T_2-T_1\right)\right)}{\sqrt{M} z_1 }\right)\,,
\end{align}
and for the AdS part of the curve
\begin{align}
    A_{(-4)}^A&=\sinh ^{-1}\left(\frac{\sqrt{1-M z_1^2}}{-P z_1}\right)\,.
\end{align}
Upon combining these expressions and using the value of the crossing point $z_1$ from \eqref{crossing2}, we obtain the total lower contribution as
\begin{align}
    &A_{(-4)}=\log \left(\frac{2 \sinh \left(\frac{1}{2} \sqrt{M} \left(T_2-T_1\right)\right)}{\epsilon \sqrt{M}}\right)+\Phi_4(T_1,T_2,M)\,.\label{A-4}
\end{align}
Where we are calling the cumbersome UV cutoff independent part as $\Phi_4(T_1, T_2,M)$ which takes the form
\begin{align}
&\Phi_4(T_1,T_2,M)=-\text{csch}^{-1}\left(\coth \left(\frac{1}{2} \sqrt{M} \left(T_1+T_2\right)\right)-\coth \left(\frac{1}{2} \sqrt{M} \left(T_2-T_1\right)\right) \text{csch}\left(\frac{1}{2} \sqrt{M} \left(T_1+T_2\right)\right)\right)\nonumber\\
&+\sinh ^{-1}\Bigg[\frac{1}{2} \left(\coth \left(\frac{\sqrt{M} T_1}{2}\right)+\coth \left(\frac{\sqrt{M} T_2}{2}\right)\right)\tanh \left(\frac{1}{2} \sqrt{M} \left(T_2-T_1\right)\right)\times\nonumber\\
&\sqrt{1-\left(\coth \left(\frac{1}{2} \sqrt{M} \left(T_1+T_2\right)\right)-\cosh \left(\frac{1}{2} \sqrt{M} \left(T_2-T_1\right)\right) \text{csch}\left(\frac{1}{2} \sqrt{M} \left(T_1+T_2\right)\right)\right)^2}\Bigg]\nonumber\,.
\end{align}
This $\Phi_4(T_1, T_2,M(\beta))$ is responsible for generating the time evolution of the holographic TEE. 

%%%%%%%%%%%%%%%%%%%%%%%%%%%%%%%%%%%%%%%%%%%%%%%%%%%%%%%%%%%%%%%%%%%%%%%%%%%
\subsubsection{Timelike segment }
%%%%%%%%%%%%%%%%%%%%%%%%%%%%%%%%%%%%%%%%%%%%%%%%%%%%%%%%%%%%%%%%%%%%%%%%%%%

Recall the discussion of the timelike curve for case 3, where the length of the timelike curve was found to be independent of the crossing point. This suggests that, in a way, the timelike curve that we obtain in this case is not distinct from the one obtained earlier in case 3. Therefore, leveraging this universality of the timelike curve, we can simply read off its length for case 4 as 
\begin{align}
    \Delta_{4}&=\frac{3\pi i}{2}\,. \label{T4}
\end{align}
\begin{figure}[h]
    \centering
 % {{\includegraphics[width=7cm]{Images/S4.pdf} }}%
   % \qquad
   {{\includegraphics[width=8cm]{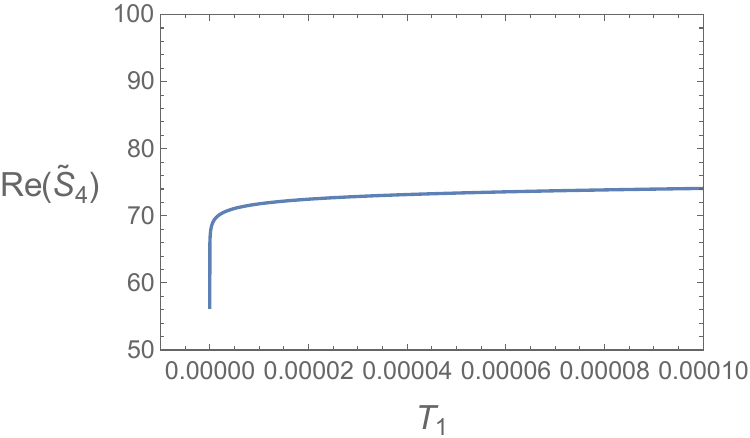} }}%
    \caption{Real part of holographic TEE for case 4. The plot uses a subregion size  $T_2-T_1=100$, $M=0.5$ and $\epsilon=0.001$.}%
    \label{Fig4}
\end{figure}

%%%%%%%%%%%%%%%%%%%%%%%%%%%%%%%%%%%%%%%%%%%%%%%%%%%%%%%%%%%%%%%%%%%%%%%%%%%
\subsubsection{Holographic TEE and hints of late time thermalization}
%%%%%%%%%%%%%%%%%%%%%%%%%%%%%%%%%%%%%%%%%%%%%%%%%%%%%%%%%%%%%%%%%%%%%%%%%%%

\begin{comment}
    \begin{figure}[!ht]
    \centering
    \includegraphics[width=0.5\linewidth]{Images/S4.pdf}
    \caption{Caption}
    \label{fig:enter-label}
\end{figure}
\end{comment}
Adding up all the individual contributions (\ref{A+4}), (\ref{A-4}) and    (\ref{T4}), we obtain
\begin{align}
  \tilde{S}_{4}&= \frac{c}{3} \log \left(\frac{\beta }{\pi \epsilon }\sinh \left(\frac{\pi}{\beta}\left(T_2-T_1\right)\right)\right) -\frac{c}{6}\Phi_4(T_1,T_2,\beta)+\frac{c\pi i}{4}\label{TEE4}\,.
\end{align}
\begin{figure}[h]
    \centering
    \includegraphics[width=0.6\linewidth]{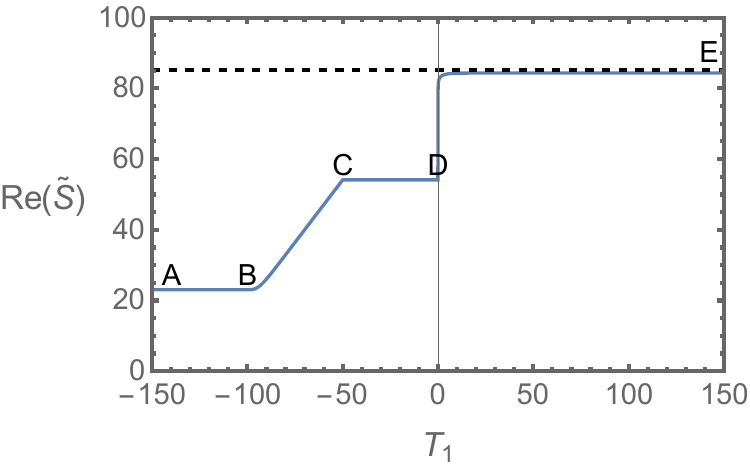}
    \caption{Combined plot for the TEE from all four cases discussed so far. The cases are named differently, and correspond to the segments AB, BC, CD and DE respectively. We have plotted it for the subregion size $T_2-T_1=100$, $M=0.5$ and $\epsilon=0.001$. The dotted black line above correspond to the TEE value for the BTZ case as described in \cite{Doi:2022iyj,Doi:2023zaf}. The DE segment asymptoting to the BTZ result is a clear indication of TEE for a thermalized CFT.}
    \label{TEE}
\end{figure}
We have plotted the real part of this in figure \ref{Fig4}. Here, we have the leading order contribution coming from the fact that the spacelike surfaces are now traversing the BTZ geometry. The fact that the lower surface is able to cross over to the pure AdS region is contained in the finite part $\Phi_4(T_1,T_2,\beta)$. Just as we remarked in the case 2, this finite piece is responsible to capture the entanglement growth as long as the crossing point is evolving along the null shell. This entanglement growth, as depicted in figure \ref{Fig4}, comprises of an initial jump from case 3 and then asymptotes precisely (with a slow growth) to the TEE value of the static BTZ \cite{Doi:2022iyj,Doi:2023zaf}! This slow growth can be attributed to the fact that the crossing point itself saturates to $z_h$.\\
%As $z_1$ approaches $z_h$, due to the crossing, the area and hence the TEE keeps growing monotonically, and it crosses the static BTZ TEE at a value $T_1=\frac{e^{\frac{\sqrt{M}T}{2}}}{2\sqrt{M}}$ (with the subregion interval length $T$ held fixed). 

%It is interesting to note that a rapid growth rate at earlier times and a slower growth rate at late times were encountered in the studies of entanglement entropy \cite{Liu:2013qca} (see also \cite{Kibe:2021qjy,Kibe:2024icu} for some related works). However, we will not quite literally draw comparisons between the two cases because our quantity of interest, as well as the entanglement structure, are quite different. The rate of increase in that case was linear, whereas the growth rate encountered here has some complicated dependence on boundary time.  In future it will be important to understand this behaviour from a purely boundary perspective.\\

At late times, once the boundary geometry has settled into a stationary configuration, the bulk geometry exterior to the event horizon undergoes dynamical relaxation and the TEE approaches the value corresponding to the final equilibrium state dual to the black hole.  In this case, the persisting growth of TEE as depicted by the DE segment in figure \ref{TEE},\footnote{In figure \ref{TEE}, we have combined the values of the TEE from all four cases.} saturates towards the black hole TEE from below. In this sense, the current measure of TEE can provide a holographic dual description that the dual field theory has undergone thermal relaxation at late times. The dotted black line in figure \ref{TEE} sets the upper bound to TEE, given by the TEE of the static BTZ case \cite{Doi:2022iyj,Doi:2023zaf}.

%%%%%%%%%%%%%%%%%%%%%%%%%%%%%%%%%%%%%%%%%%%%%%%%%%%%%%%%%%%%%%%%%%%%%%%%%%%
\section{Conclusions and outlook}\label{sec:conclude}
%%%%%%%%%%%%%%%%%%%%%%%%%%%%%%%%%%%%%%%%%%%%%%%%%%%%%%%%%%%%%%%%%%%%%%%%%%%

In this paper, we have taken up a holographic study of the recently developed measure of timelike entanglement for temporal subregions within the framework of Lorentzian AdS/CFT. Even though it is not quite clear to what extent the TEE measures statistical properties of the states, it has a close connection with the pseudo-entropy developed in the context of non-hermitian states in the Euclidean CFT dual to de Sitter. One of the aspects that we found particularly interesting has to do with the dual geometries containing a black hole. In that case, the holographic extremal surfaces measuring the TEE quite generically see the inside-horizon region. This feature may then allow us to study BH physics more effectively via TEE. As mentioned in the introduction, already a vast amount of literature exists that models such a process in terms of a dynamical AdS Vaidya geometry. A study of spacelike entanglement in those cases indeed shows expected behaviour during thermalization. It was therefore a natural question for us to investigate to what extent (if at all) these processes are imprinted in TEE.\\
%The particular question that we focused on here is a holographic study of how a CFT thermalizes after undergoing a global quench.

And the features that we have uncovered here seem encouraging and compelling. Modelling the CFT quench in terms of AdS Vaidya, we investigated if TEE\footnote{Of course, when we say TEE, we have in mind the (real) area of the spacelike green segments of the extremal surface, as in figure \ref{AdSBTZ}.} changes suggestively as the boundary interval probes various stages of the CFT's life. At its infancy (case 1), we recovered the known AdS results for TEE, as was obtained in \cite{Doi:2022iyj,Doi:2023zaf}. However, as the interval started to become aware of the shock (particularly in cases 2 and 4), it started behaving dynamically. In particular, we found that as soon as the spacelike parts of the extremal hypersurfaces make a journey from BH to AdS, TEE starts growing linearly with time. For example, keeping the interval size fixed, when we varied one of the endpoints of the interval, it showed a clear growth in cases 2 and 4 (figures \ref{TEE2} and \ref{Fig4} respectively). The TEE first shows a linear rise as long as the upper surface straddles across the shell in case 2. Also there is a growth for a short time in case 4 when the lower surface straddles across the shell, before it settles down to the asymptotic upper bound set by the TEE of the static BTZ. Growths of different rates is also a property of HEE \cite{Liu:2013qca}, which increases rather rapidly (quadratically in time) in the earlier stages of thermalization and thermalizes at very late times \cite{Calabrese:2005in,Balasubramanian:2011ur,Liu:2013iza,Liu:2013qca}. However, as remarked earlier, we will rather not push this analogy too far because the entanglement measure under consideration is rather different. Only after we find a better handle on CFT thermalization in terms of TEE, can we meaningfully draw comparisons between the two cases.  In another interesting note, we found that the spacelike surfaces crossed the shell in such a clever way, so as to dynamically impose causal natures of the associated timelike segment. For us this was manifested by the saturation of $z_1$ to a horizon value $z_h$ (see figure \ref{z2}).
\\

Finally, we are left with more questions than answers. Here, we have merely supplied some tools to facilitate a more thorough study of thermalization in the future investigations, but we have not explored the mechanism in full detail. This is because dealing with TEE for timelike intervals comes with its own complications. In particular, our case 2 is quite unique and doesn't appear in any prior scientific studies, in which the entangling region itself straddles the shell. What we found is that the junctions conditions shows some inconsistencies in this case (appendix \ref{app:case2alt}), and we had to resort to the gluing conditions as mentioned in section \ref{subsec:straddleearly}. This requires further studies. In addition, most of the studies of TEE seem to be primarily restricted in (2+1) dimensional bulk, because it is operationally easier to compute the lengths of the extremal curves in  AdS$_3$/CFT$_2$. We have seen that the equations of motion only supply us with the solutions for spacelike geodesics, and the adjoining timelike geodesics remain elusive. It will be very interesting to see how the contributions from the timelike geodesics, and ultimately the entire TEE computations can be obtained in higher dimensions.
%\DS{Added some lines here} 

%%%%%%%%%%%%%%%%%%%%%%%%%%%%%%%%%%%%%%%%%%%%%%%%%%%%%%%%%%%%%%%%%%%%%%%%%%%
\paragraph{Acknowledgements}
%%%%%%%%%%%%%%%%%%%%%%%%%%%%%%%%%%%%%%%%%%%%%%%%%%%%%%%%%%%%%%%%%%%%%%%%%%%

$\newline$
\\
We would like to thank Ayan Mukhopadhyay and Shubho Roy for fruitful discussions. BS thanks CSIR-HRDG for financial support with fellowship number 09/1022(12312)/2021-EMR-I. The work of GK is supported by SERB sponsored Research Project no. CRG/2023/000904.  The work of DS is supported by the DST-FIST grant number SR/FST/PSI-
225/2016, SERB MATRICS grant MTR/2021/000168 and SERB CRG grant CRG/2023/000904. This work was done in part during the workshop ``Holographic Duality and Models of Quantum Computation'' held at Tsinghua Southeast Asia Center on Bali, Indonesia (2024).

%%%%%%%%%%%%%%%%%%%%%%%%%%%%%%%%%%%%%%%%%%%%%%%%%%%%%%%%%%%%%%%%%%%%%%%%%%%
\appendix
%%%%%%%%%%%%%%%%%%%%%%%%%%%%%%%%%%%%%%%%%%%%%%%%%%%%%%%%%%%%%%%%%%%%%%%%%%%

%%%%%%%%%%%%%%%%%%%%%%%%%%%%%%%%%%%%%%%%%%%%%%%%%%%%%%%%%%%%%%%%%%%%%%%%%%%
\section{Variational treatment for the pure AdS case \label{var}}
%%%%%%%%%%%%%%%%%%%%%%%%%%%%%%%%%%%%%%%%%%%%%%%%%%%%%%%%%%%%%%%%%%%%%%%%%%%

The domains of validity of the integration constants are $c_1>T_1$ and $c_2<T_2$. In this case the total area of the spacelike curves is (adding \eqref{AdS-1} and \eqref{AdS+1})
\begin{align}
A_{AdS}=A_{(-1)}+A_{(+1)}&=   \sinh ^{-1}\left(\frac{(T_2-c_2)}{\epsilon }\right)-\sinh ^{-1}\left(\frac{(T_1-c_1)}{\epsilon }\right)\,.\nonumber
\end{align}
From the conservation of momentum we have
\begin{align}\label{eq:cconstraint}
    T_2-c_2=c_1-T_1\,,
\end{align}
using which we can write
\begin{align}
    A_{AdS}&=  2 \sinh ^{-1}\left(\frac{T_2-c_2}{\epsilon }\right)\,.\nonumber
\end{align}
We can then expect to obtain the correct area by demanding an extremization procedure with respect to $c_2$. This gives
\begin{align}
   \delta_{c_2} {A}_{AdS}=\left(\frac{2}{\epsilon  \sqrt{\frac{\left(T_2-c_2\right){}^2}{\epsilon ^2}+1}}\right)\delta c_2=0\nonumber.
\end{align}
As we can see, the extremization process does not fix $c_2$, and hence we are free to choose (keeping in mind \eqref{eq:cconstraint}),
\begin{align}
c_1=c_2=\frac{T_1+T_2}{2}\,.\nonumber
\end{align}
This is indeed the most optimal choice. Not only that it is symmetric in $T_1$ and $T_2$, it also makes the final area to depend only on the interval size and reproduces the known results of \cite{Doi:2022iyj,Doi:2023zaf}
\begin{align}
   {A}_{AdS}&=2 \sinh ^{-1}\left(\frac{T_1-T_2}{2 \epsilon }\right)\,.\nonumber
\end{align}
As noted during the main text, we will keep the same choices for the constants (as long as applicable) for a smooth transition of extremal area surfaces from one case to another.

%%%%%%%%%%%%%%%%%%%%%%%%%%%%%%%%%%%%%%%%%%%%%%%%%%%%%%%%%%%%%%%%%%%%%%%%%%%
\section{Junction conditions \label{app:jc}}
%%%%%%%%%%%%%%%%%%%%%%%%%%%%%%%%%%%%%%%%%%%%%%%%%%%%%%%%%%%%%%%%%%%%%%%%%%%

The Euler-Lagrange equations coming from the Lagrangian in the action (\ref{Action}) are
\begin{align}
    \frac{d}{dz}\left(\frac{f(z,v)v'(z)+1}{z\sqrt{Q}}\right)&=\frac{v'^2}{2 z\sqrt{Q}}\frac{\partial f(z,v)}{\partial v}\,,\label{veom}
\end{align}
and 
\begin{align}
   \frac{x'(z)}{z \sqrt{Q}}=J\,,\label{xeom}
\end{align}
where we have defined $Q\equiv -f(z) v'(z)^2-2 v'(z)-x'(z)^2$. Below we will rewrite (\ref{veom}) as 
\begin{align}
    \frac{\partial }{\partial z}\left(\frac{f(z,v)v'(z)+1}{z\sqrt{Q}}\right)+v'\frac{\partial }{\partial v}\left(\frac{f(z,v)v'(z)+1}{z\sqrt{Q}}\right)&=\frac{v'^2}{2 z\sqrt{Q}}\frac{\partial f(z,v)}{\partial v}\,.\nonumber
\end{align}
While integrating along the $v$ direction (across the thin shell, and holding $z$ fixed), we can drop the first term of the above. Subsequently, using (\ref{xeom}), we can write 
\begin{align}
   \frac{\partial }{\partial v}\left(\frac{(f(z,v)v'(z)+1)}{x'}\right)&=\frac{v'}{2x'}\frac{\partial f(z,v)}{\partial v}\,.\nonumber
\end{align}
%and $\frac{\partial v'}{\partial v}=0$
Since $x'$ has no dependence upon $v$, we can safely write 
\begin{align}
   %\frac{\partial }{\partial v}\left(v' f \right)&=\frac{1}{2}\frac{\partial (v'f)}{\partial v}.\nonumber
 \frac{1}{2 x'}  \frac{\partial }{\partial v}\left(v' f \right)=0\,.\nonumber
\end{align} 
Integrating both sides with respect to $v$ over an infinitesimal interval across the shell (with fixed $z$), we get (once again, superscripts $A$ and $B$ signifying AdS and BH regions)
\begin{align}
  f^B(z_c)v'^B(z_c)-  f^A(z_c)v'^A(z_c)=0\,.\label{junction}
\end{align}
This is the resulting joining condition across the shell, which we will use when applicable. Below, we also provide an alternative derivation of the same.\\

We can also follow the treatment of \cite{Balasubramanian:2011ur}, where we consider a point on the shell and take the coordinate differences between this point and the inside and outside points on either side of the shell to be $\Delta X_i=(\Delta z_i,\Delta v_i,\Delta x_i)$ and $\Delta X_o=(\Delta z_o,\Delta v_o,\Delta x_o)$ respectively. Subscripts $i$ and $o$ here stand for inside and outside. If the coordinate differences between the points themselves are $\Delta X=(\Delta z,\Delta v,\Delta x)$, such that $\Delta X=\Delta X_i+
\Delta X_o$, then the distance between the points outside and inside via the point on the shell will be 
\begin{align}
    \Delta s&=\sqrt{\Delta s_i^2}+\sqrt{\Delta s_o^2}
    \nonumber\\
    &=\sqrt{\frac{1}{z_i^2}\left(-f_i\Delta v_i^2-2\Delta v_i\Delta z_i+\Delta x_i^2\right)}+\sqrt{\frac{1}{z_o^2}\left(-f_o\Delta v_o^2-2\Delta v_o\Delta z_o+\Delta x_o^2\right)}\,.\nonumber
\end{align}
Upon extremizing with respect to $\Delta z_i$ and $\Delta x_i$, and using $\frac{\delta \Delta X_o}{\delta \Delta X_i}=-1$ leads us to  
\begin{align}
   \frac{\Delta v_i}{z_i^2\sqrt{\Delta s_i^2}}=\frac{\Delta v_o}{z_o^2\sqrt{\Delta s_o^2}}\label{varz}\,,
\end{align}
and
%Upon extremizing with respect to $\Delta x_i$
%and using $\frac{\delta \Delta x_o}{\delta \Delta x_i}=-1$  leads us to  
\begin{align}
   \frac{\Delta x_i}{z_i^2\sqrt{\Delta s_i^2}}=\frac{\Delta x_o}{z_o^2\sqrt{\Delta s_o^2}}\,.\label{varx}
\end{align}
Upon combining them, we get 
\begin{align}
    \frac{\Delta x_i}{\Delta v_i}&= \frac{\Delta x_o}{\Delta v_o}\,.\label{varxv} 
\end{align}
Similarly, taking variation with respect to $\Delta v_i$  and then extremizing leads to
\begin{align}
    \frac{\Delta v_i f_i+\Delta z_i}{z_i^2\sqrt{\Delta s_i^2 }}= \frac{\Delta v_o f_o+\Delta z_o}{z_o^2\sqrt{\Delta s_o^2 }}\,.\nonumber
\end{align}
Combining with (\ref{varz}) to eliminate extra factors, gives 
\begin{align}
  f_i +    \frac{\Delta z_i}{\Delta v_i}&= f_o +    \frac{\Delta z_o}{\Delta v_o}\,.\nonumber
  \end{align}
%  From (\ref{varxv}), we may eliminate the $x$-part so that this simplifies to 
The above equation can be massaged to rewrite in the following form
  \begin{align}
  \frac{\Delta z_i}{\Delta v_i}\left( f_i  \frac{\Delta v_i }{\Delta z_i}+    1\right)&=\frac{\Delta z_o}{\Delta v_o}\left( f_o  \frac{\Delta v_o }{\Delta z_o}+    1\right)\,,\nonumber\\
  \frac{\Delta v_o}{\Delta v_i}\left( f_i  \frac{\Delta v_i }{\Delta z_i}+    1\right)&=\frac{\Delta z_o}{\Delta z_i}\left( f_o  \frac{\Delta v_o }{\Delta z_o}+    1\right)\,,\nonumber\\
 \left( f_i  \frac{\Delta v_i }{\Delta z_i}+    1\right)&= \left( f_o  \frac{\Delta v_o }{\Delta z_o}+    1\right)\,.\nonumber
\end{align}
%\begin{align}
 % f_i  \frac{\Delta v_i }{\Delta x_i}+    \frac{\Delta z_i}{\Delta x_i}&= f_o  \frac{\Delta v_o }{\Delta x_o}+    \frac{\Delta z_o}{\Delta x_o}\nonumber,\\
  %\frac{\Delta z_i}{\Delta x_i}\left( f_i  \frac{\Delta v_i }{\Delta x_i}\frac{\Delta x_i}{\Delta z_i}+    1\right)&=\frac{\Delta z_o}{\Delta x_o}\left( f_o  \frac{\Delta v_o }{\Delta x_o}\frac{\Delta x_o}{\Delta z_o}+    1\right)\nonumber,\\
 % \frac{\Delta x_o}{\Delta x_i}\left( f_i  \frac{\Delta v_i }{\Delta z_i}+    1\right)&=\frac{\Delta z_o}{\Delta z_i}\left( f_o  \frac{\Delta v_o }{\Delta z_o}+    1\right)\nonumber,\\
 %\left( f_i  \frac{\Delta v_i }{\Delta z_i}+    1\right)&= \left( f_o  \frac{\Delta v_o }{\Delta z_o}+    1\right)\nonumber.
%\end{align}
In the limit as $\Delta X \to 0$, we recover (\ref{junction}).

%%%%%%%%%%%%%%%%%%%%%%%%%%%%%%%%%%%%%%%%%%%%%%%%%%%%%%%%%%%%%%%%%%%%%%%%%%%

%%%%%%%%%%%%%%%%%%%%%%%%%%%%%%%%%%%%%%%%%%%%%%%%%%%%%%%%%%%%%%%%%%%%%%%%%%%
\section{Subtleties with case 2 \label{app:case2alt}}
By using the junction conditions \eqref{eq:jc}, we can write down
\begin{align}
    \frac{1-M z_2^2}{z_2^2 \left(P^2-M\right)+\sqrt{P^2 z_2^2 \left(z_2^2 \left(P^2-M\right)+1\right)}+1}=\frac{1}{q^2 z_2^2+\sqrt{q^2 z_2^2 \left(q^2 z_2^2+1\right)}+1}\,.
\end{align}
This relates the first integrals on either side of the shell via (compare with how it went in case 4, as discussed in section \ref{subsec:afterqq})
\begin{equation}
	q=\pm \frac{P}{\sqrt{1-Mz_2^2}}\,.
\end{equation}
However, because in this case 2, $z_2$ goes all the way to infinity, it turns out that this function is not well-behaved over the entire range of $z_2$. In particular, the above equation has a solution when $1-Mz_2^2>0$ but not for $1-Mz_2^2<0$. Therefore, the junction condition doesn't appear to be applicable for the case 2. One can in principle go ahead with a similar computation as in case 4, but it yields unphysical results such as decrease in the real part of TEE with time. 

%%%%%%%%%%%%%%%%%%%%%%%%%%%%%%%%%%%%%%%%%%%%%%%%%%%%%%%%%%%%%%%%%%%%%%%%%%%

\bibliographystyle{utphys}
\bibliography{ref}

\providecommand{\href}[2]{#2}\begingroup\raggedright\begin{thebibliography}{10}

\bibitem{Maldacena:1997re}
J.~M. Maldacena, ``{The Large N limit of superconformal field theories and supergravity},'' \href{https://dx.doi.org/10.1023/A:1026654312961}{{\em Int. J. Theor. Phys.} {\bfseries 38} (1999) 1113--1133}, \href{https://arxiv.org/abs/hep-th/9711200}{{\ttfamily arXiv:hep-th/9711200}}.

\bibitem{Gubser:1998bc}
S.~Gubser, I.~R. Klebanov, and A.~M. Polyakov, ``{Gauge theory correlators from noncritical string theory},'' \href{https://dx.doi.org/10.1016/S0370-2693(98)00377-3}{{\em Phys. Lett. B} {\bfseries 428} (1998) 105--114}, \href{https://arxiv.org/abs/hep-th/9802109}{{\ttfamily arXiv:hep-th/9802109}}.

\bibitem{Witten:1998qj}
E.~Witten, ``{Anti-de Sitter space and holography},'' \href{https://dx.doi.org/10.4310/ATMP.1998.v2.n2.a2}{{\em Adv. Theor. Math. Phys.} {\bfseries 2} (1998) 253--291}, \href{https://arxiv.org/abs/hep-th/9802150}{{\ttfamily arXiv:hep-th/9802150}}.

\bibitem{Banados:1992wn}
M.~Banados, C.~Teitelboim, and J.~Zanelli, ``{The Black hole in three-dimensional space-time},'' \href{https://dx.doi.org/10.1103/PhysRevLett.69.1849}{{\em Phys. Rev. Lett.} {\bfseries 69} (1992) 1849--1851}, \href{https://arxiv.org/abs/hep-th/9204099}{{\ttfamily arXiv:hep-th/9204099}}.

\bibitem{Abajo-Arrastia:2010ajo}
J.~Abajo-Arrastia, J.~Aparicio, and E.~Lopez, ``{Holographic Evolution of Entanglement Entropy},'' \href{https://dx.doi.org/10.1007/JHEP11(2010)149}{{\em JHEP} {\bfseries 11} (2010) 149}, \href{https://arxiv.org/abs/1006.4090}{{\ttfamily arXiv:1006.4090 [hep-th]}}.

\bibitem{Ebrahim:2010ra}
H.~Ebrahim and M.~Headrick, ``{Instantaneous Thermalization in Holographic Plasmas},'' \href{https://arxiv.org/abs/1010.5443}{{\ttfamily arXiv:1010.5443 [hep-th]}}.

\bibitem{Albash:2010mv}
T.~Albash and C.~V. Johnson, ``{Evolution of Holographic Entanglement Entropy after Thermal and Electromagnetic Quenches},'' \href{https://dx.doi.org/10.1088/1367-2630/13/4/045017}{{\em New J. Phys.} {\bfseries 13} (2011) 045017}, \href{https://arxiv.org/abs/1008.3027}{{\ttfamily arXiv:1008.3027 [hep-th]}}.

\bibitem{Balasubramanian:2010ce}
V.~Balasubramanian, A.~Bernamonti, J.~de~Boer, N.~Copland, B.~Craps, E.~Keski-Vakkuri, B.~Muller, A.~Schafer, M.~Shigemori, and W.~Staessens, ``{Thermalization of Strongly Coupled Field Theories},'' \href{https://dx.doi.org/10.1103/PhysRevLett.106.191601}{{\em Phys. Rev. Lett.} {\bfseries 106} (2011) 191601}, \href{https://arxiv.org/abs/1012.4753}{{\ttfamily arXiv:1012.4753 [hep-th]}}.

\bibitem{Balasubramanian:2011ur}
V.~Balasubramanian, A.~Bernamonti, J.~de~Boer, N.~Copland, B.~Craps, E.~Keski-Vakkuri, B.~Muller, A.~Schafer, M.~Shigemori, and W.~Staessens, ``{Holographic Thermalization},'' \href{https://dx.doi.org/10.1103/PhysRevD.84.026010}{{\em Phys. Rev. D} {\bfseries 84} (2011) 026010}, \href{https://arxiv.org/abs/1103.2683}{{\ttfamily arXiv:1103.2683 [hep-th]}}.

\bibitem{Caceres:2012em}
E.~Caceres and A.~Kundu, ``{Holographic Thermalization with Chemical Potential},'' \href{https://dx.doi.org/10.1007/JHEP09(2012)055}{{\em JHEP} {\bfseries 09} (2012) 055}, \href{https://arxiv.org/abs/1205.2354}{{\ttfamily arXiv:1205.2354 [hep-th]}}.

\bibitem{Baron:2012fv}
W.~Baron, D.~Galante, and M.~Schvellinger, ``{Dynamics of holographic thermalization},'' \href{https://dx.doi.org/10.1007/JHEP03(2013)070}{{\em JHEP} {\bfseries 03} (2013) 070}, \href{https://arxiv.org/abs/1212.5234}{{\ttfamily arXiv:1212.5234 [hep-th]}}.

\bibitem{Arefeva:2013kvb}
I.~Aref'eva, A.~Bagrov, and A.~S. Koshelev, ``{Holographic Thermalization from Kerr-AdS},'' \href{https://dx.doi.org/10.1007/JHEP07(2013)170}{{\em JHEP} {\bfseries 07} (2013) 170}, \href{https://arxiv.org/abs/1305.3267}{{\ttfamily arXiv:1305.3267 [hep-th]}}.

\bibitem{Liu:2013qca}
H.~Liu and S.~J. Suh, ``{Entanglement growth during thermalization in holographic systems},'' \href{https://dx.doi.org/10.1103/PhysRevD.89.066012}{{\em Phys. Rev. D} {\bfseries 89} no.~6, (2014) 066012}, \href{https://arxiv.org/abs/1311.1200}{{\ttfamily arXiv:1311.1200 [hep-th]}}.

\bibitem{Liu:2013iza}
H.~Liu and S.~J. Suh, ``{Entanglement Tsunami: Universal Scaling in Holographic Thermalization},'' \href{https://dx.doi.org/10.1103/PhysRevLett.112.011601}{{\em Phys. Rev. Lett.} {\bfseries 112} (2014) 011601}, \href{https://arxiv.org/abs/1305.7244}{{\ttfamily arXiv:1305.7244 [hep-th]}}.

\bibitem{Ryu:2006bv}
S.~Ryu and T.~Takayanagi, ``{Holographic derivation of entanglement entropy from AdS/CFT},'' \href{https://dx.doi.org/10.1103/PhysRevLett.96.181602}{{\em Phys. Rev. Lett.} {\bfseries 96} (2006) 181602}, \href{https://arxiv.org/abs/hep-th/0603001}{{\ttfamily arXiv:hep-th/0603001}}.

\bibitem{Hubeny:2007xt}
V.~E. Hubeny, M.~Rangamani, and T.~Takayanagi, ``{A Covariant holographic entanglement entropy proposal},'' \href{https://dx.doi.org/10.1088/1126-6708/2007/07/062}{{\em JHEP} {\bfseries 07} (2007) 062}, \href{https://arxiv.org/abs/0705.0016}{{\ttfamily arXiv:0705.0016 [hep-th]}}.

\bibitem{Hubeny:2012ry}
V.~E. Hubeny, ``{Extremal surfaces as bulk probes in AdS/CFT},'' \href{https://dx.doi.org/10.1007/JHEP07(2012)093}{{\em JHEP} {\bfseries 07} (2012) 093}, \href{https://arxiv.org/abs/1203.1044}{{\ttfamily arXiv:1203.1044 [hep-th]}}.

\bibitem{Hartman:2013qma}
T.~Hartman and J.~Maldacena, ``{Time Evolution of Entanglement Entropy from Black Hole Interiors},'' \href{https://dx.doi.org/10.1007/JHEP05(2013)014}{{\em JHEP} {\bfseries 05} (2013) 014}, \href{https://arxiv.org/abs/1303.1080}{{\ttfamily arXiv:1303.1080 [hep-th]}}.

\bibitem{Freivogel:2014lja}
B.~Freivogel, R.~Jefferson, L.~Kabir, B.~Mosk, and I.-S. Yang, ``{Casting Shadows on Holographic Reconstruction},'' \href{https://dx.doi.org/10.1103/PhysRevD.91.086013}{{\em Phys. Rev. D} {\bfseries 91} no.~8, (2015) 086013}, \href{https://arxiv.org/abs/1412.5175}{{\ttfamily arXiv:1412.5175 [hep-th]}}.

\bibitem{Buchel:2014gta}
A.~Buchel, R.~C. Myers, and A.~van Niekerk, ``{Nonlocal probes of thermalization in holographic quenches with spectral methods},'' \href{https://dx.doi.org/10.1007/JHEP02(2015)017}{{\em JHEP} {\bfseries 02} (2015) 017}, \href{https://arxiv.org/abs/1410.6201}{{\ttfamily arXiv:1410.6201 [hep-th]}}. [Erratum: JHEP 07, 137 (2015)].

\bibitem{Aparicio:2011zy}
J.~Aparicio and E.~Lopez, ``{Evolution of Two-Point Functions from Holography},'' \href{https://dx.doi.org/10.1007/JHEP12(2011)082}{{\em JHEP} {\bfseries 12} (2011) 082}, \href{https://arxiv.org/abs/1109.3571}{{\ttfamily arXiv:1109.3571 [hep-th]}}.

\bibitem{Narayan:2015vda}
K.~Narayan, ``{Extremal surfaces in de Sitter spacetime},'' \href{https://dx.doi.org/10.1103/PhysRevD.91.126011}{{\em Phys. Rev. D} {\bfseries 91} no.~12, (2015) 126011}, \href{https://arxiv.org/abs/1501.03019}{{\ttfamily arXiv:1501.03019 [hep-th]}}.

\bibitem{Narayan:2016xwq}
K.~Narayan, ``{On $dS_4$ extremal surfaces and entanglement entropy in some ghost CFTs},'' \href{https://dx.doi.org/10.1103/PhysRevD.94.046001}{{\em Phys. Rev. D} {\bfseries 94} no.~4, (2016) 046001}, \href{https://arxiv.org/abs/1602.06505}{{\ttfamily arXiv:1602.06505 [hep-th]}}.

\bibitem{Jatkar:2017jwz}
D.~P. Jatkar and K.~Narayan, ``{Ghost-spin chains, entanglement and $bc$-ghost CFTs},'' \href{https://dx.doi.org/10.1103/PhysRevD.96.106015}{{\em Phys. Rev. D} {\bfseries 96} no.~10, (2017) 106015}, \href{https://arxiv.org/abs/1706.06828}{{\ttfamily arXiv:1706.06828 [hep-th]}}.

\bibitem{Narayan:2017xca}
K.~Narayan, ``{On extremal surfaces and de Sitter entropy},'' \href{https://dx.doi.org/10.1016/j.physletb.2018.02.010}{{\em Phys. Lett. B} {\bfseries 779} (2018) 214--222}, \href{https://arxiv.org/abs/1711.01107}{{\ttfamily arXiv:1711.01107 [hep-th]}}.

\bibitem{Mollabashi:2020yie}
A.~Mollabashi, N.~Shiba, T.~Takayanagi, K.~Tamaoka, and Z.~Wei, ``{Pseudo Entropy in Free Quantum Field Theories},'' \href{https://dx.doi.org/10.1103/PhysRevLett.126.081601}{{\em Phys. Rev. Lett.} {\bfseries 126} no.~8, (2021) 081601}, \href{https://arxiv.org/abs/2011.09648}{{\ttfamily arXiv:2011.09648 [hep-th]}}.

\bibitem{Nishioka:2021cxe}
T.~Nishioka, T.~Takayanagi, and Y.~Taki, ``{Topological pseudo entropy},'' \href{https://dx.doi.org/10.1007/JHEP09(2021)015}{{\em JHEP} {\bfseries 09} (2021) 015}, \href{https://arxiv.org/abs/2107.01797}{{\ttfamily arXiv:2107.01797 [hep-th]}}.

\bibitem{Doi:2022iyj}
K.~Doi, J.~Harper, A.~Mollabashi, T.~Takayanagi, and Y.~Taki, ``{Pseudoentropy in dS/CFT and Timelike Entanglement Entropy},'' \href{https://dx.doi.org/10.1103/PhysRevLett.130.031601}{{\em Phys. Rev. Lett.} {\bfseries 130} no.~3, (2023) 031601}, \href{https://arxiv.org/abs/2210.09457}{{\ttfamily arXiv:2210.09457 [hep-th]}}.

\bibitem{Narayan:2022afv}
K.~Narayan, ``{de Sitter space, extremal surfaces, and time entanglement},'' \href{https://dx.doi.org/10.1103/PhysRevD.107.126004}{{\em Phys. Rev. D} {\bfseries 107} no.~12, (2023) 126004}, \href{https://arxiv.org/abs/2210.12963}{{\ttfamily arXiv:2210.12963 [hep-th]}}.

\bibitem{Doi:2023zaf}
K.~Doi, J.~Harper, A.~Mollabashi, T.~Takayanagi, and Y.~Taki, ``{Timelike entanglement entropy},'' \href{https://dx.doi.org/10.1007/JHEP05(2023)052}{{\em JHEP} {\bfseries 05} (2023) 052}, \href{https://arxiv.org/abs/2302.11695}{{\ttfamily arXiv:2302.11695 [hep-th]}}.

\bibitem{Narayan:2023ebn}
K.~Narayan and H.~K. Saini, ``{Notes on time entanglement and pseudo-entropy},'' \href{https://dx.doi.org/10.1140/epjc/s10052-024-12855-x}{{\em Eur. Phys. J. C} {\bfseries 84} no.~5, (2024) 499}, \href{https://arxiv.org/abs/2303.01307}{{\ttfamily arXiv:2303.01307 [hep-th]}}.

\bibitem{Parzygnat:2023avh}
A.~J. Parzygnat, T.~Takayanagi, Y.~Taki, and Z.~Wei, ``{SVD entanglement entropy},'' \href{https://dx.doi.org/10.1007/JHEP12(2023)123}{{\em JHEP} {\bfseries 12} (2023) 123}, \href{https://arxiv.org/abs/2307.06531}{{\ttfamily arXiv:2307.06531 [hep-th]}}.

\bibitem{Narayan:2023zen}
K.~Narayan, ``{Further remarks on de Sitter space, extremal surfaces, and time entanglement},'' \href{https://dx.doi.org/10.1103/PhysRevD.109.086009}{{\em Phys. Rev. D} {\bfseries 109} no.~8, (2024) 086009}, \href{https://arxiv.org/abs/2310.00320}{{\ttfamily arXiv:2310.00320 [hep-th]}}.

\bibitem{Goswami:2024vfl}
K.~Goswami, K.~Narayan, and G.~Yadav, ``{No-boundary extremal surfaces in slow-roll inflation and other cosmologies},'' \href{https://dx.doi.org/10.1007/JHEP03(2025)193}{{\em JHEP} {\bfseries 03} (2025) 193}, \href{https://arxiv.org/abs/2409.14208}{{\ttfamily arXiv:2409.14208 [hep-th]}}.

\bibitem{Strominger:2001pn}
A.~Strominger, ``{The dS / CFT correspondence},'' \href{https://dx.doi.org/10.1088/1126-6708/2001/10/034}{{\em JHEP} {\bfseries 10} (2001) 034}, \href{https://arxiv.org/abs/hep-th/0106113}{{\ttfamily arXiv:hep-th/0106113}}.

\bibitem{Das:2023yyl}
A.~Das, S.~Sachdeva, and D.~Sarkar, ``{Bulk reconstruction using timelike entanglement in (A)dS},'' \href{https://dx.doi.org/10.1103/PhysRevD.109.066007}{{\em Phys. Rev. D} {\bfseries 109} no.~6, (2024) 066007}, \href{https://arxiv.org/abs/2312.16056}{{\ttfamily arXiv:2312.16056 [hep-th]}}.

\bibitem{Anegawa:2024kdj}
T.~Anegawa and K.~Tamaoka, ``{Black hole singularity and timelike entanglement},'' \href{https://dx.doi.org/10.1007/JHEP10(2024)182}{{\em JHEP} {\bfseries 10} (2024) 182}, \href{https://arxiv.org/abs/2406.10968}{{\ttfamily arXiv:2406.10968 [hep-th]}}.

\bibitem{Heller:2024whi}
M.~P. Heller, F.~Ori, and A.~Serantes, ``{Geometric interpretation of timelike entanglement entropy},'' \href{https://arxiv.org/abs/2408.15752}{{\ttfamily arXiv:2408.15752 [hep-th]}}.

\bibitem{Jiang:2023loq}
X.~Jiang, P.~Wang, H.~Wu, and H.~Yang, ``{Timelike entanglement entropy in dS$_{3}$/CFT$_{2}$},'' \href{https://dx.doi.org/10.1007/JHEP08(2023)216}{{\em JHEP} {\bfseries 08} (2023) 216}, \href{https://arxiv.org/abs/2304.10376}{{\ttfamily arXiv:2304.10376 [hep-th]}}.

\bibitem{Balasubramanian:2012tu}
V.~Balasubramanian, A.~Bernamonti, B.~Craps, V.~Ker\"anen, E.~Keski-Vakkuri, B.~M\"uller, L.~Thorlacius, and J.~Vanhoof, ``{Thermalization of the spectral function in strongly coupled two dimensional conformal field theories},'' \href{https://dx.doi.org/10.1007/JHEP04(2013)069}{{\em JHEP} {\bfseries 04} (2013) 069}, \href{https://arxiv.org/abs/1212.6066}{{\ttfamily arXiv:1212.6066 [hep-th]}}.

\bibitem{He:2024emd}
J.-H. He and R.-Q. Yang, ``{Geodesics connecting endpoints of timelike interval in an asymptotically AdS spacetime},'' \href{https://dx.doi.org/10.1103/PhysRevD.111.026024}{{\em Phys. Rev. D} {\bfseries 111} no.~2, (2025) 026024}, \href{https://arxiv.org/abs/2408.04783}{{\ttfamily arXiv:2408.04783 [hep-th]}}.

\bibitem{Heller:2025kvp}
M.~P. Heller, F.~Ori, and A.~Serantes, ``{Temporal Entanglement from Holographic Entanglement Entropy},'' \href{https://arxiv.org/abs/2507.17847}{{\ttfamily arXiv:2507.17847 [hep-th]}}.

\bibitem{Nunez:2025ppd}
C.~Nunez and D.~Roychowdhury, ``{Interpolating between Space-like and Time-like Entanglement via Holography},'' \href{https://arxiv.org/abs/2507.17805}{{\ttfamily arXiv:2507.17805 [hep-th]}}.

\bibitem{messiah1999quantum}
A.~Messiah, {\em Quantum Mechanics}.
\newblock Dover books on physics. Dover Publications, 1999.
\newblock \url{https://books.google.co.in/books?id=mwssSDXzkNcC}.

\bibitem{Hamilton:2006fh}
A.~Hamilton, D.~N. Kabat, G.~Lifschytz, and D.~A. Lowe, ``{Local bulk operators in AdS/CFT: A Holographic description of the black hole interior},'' \href{https://dx.doi.org/10.1103/PhysRevD.75.106001}{{\em Phys. Rev. D} {\bfseries 75} (2007) 106001}, \href{https://arxiv.org/abs/hep-th/0612053}{{\ttfamily arXiv:hep-th/0612053}}. [Erratum: Phys.Rev.D 75, 129902 (2007)].

\bibitem{Calabrese:2005in}
P.~Calabrese and J.~L. Cardy, ``{Evolution of entanglement entropy in one-dimensional systems},'' \href{https://dx.doi.org/10.1088/1742-5468/2005/04/P04010}{{\em J. Stat. Mech.} {\bfseries 0504} (2005) P04010}, \href{https://arxiv.org/abs/cond-mat/0503393}{{\ttfamily arXiv:cond-mat/0503393}}.

\end{thebibliography}\endgroup
\end{document}